%% file: FBAlAtt_Dec13.tex
\documentclass[11pt,onecolumn,twoside,notitlepage,fleqn,letterpaper,final]{article}[]
\usepackage{amsmath}
\pdfoutput=1
\input{texsetup.tex}

\newcommand{\gev}{\mrm{\:GeV}}
\newcommand{\tev}{\mrm{\:TeV}}
\newcommand{\ttbar}{{t\bar t}}
\newcommand{\pTt}{{p_{T,\ttbar}}}
\newcommand{\mtt}{{m_{t\bar t}}}
\newcommand{\Al}{{A_{l}}}
\newcommand{\Alm}{{A_{l}\,(\mtt)}}
\newcommand{\Alp}{{A_{l}\,(p_T^l)}}
\newcommand{\Alpb}{{A^\mrm{boost}_l\,(p_T^l)}}
\newcommand{\All}{{A_{ll}}}
\newcommand{\At}{{A_{\ttbar}}}
\newcommand{\Atm}{{A_{\ttbar}\,(\mtt)}}
\newcommand{\Atp}{{A_{\ttbar}}\,(p_T^l)}
\newcommand{\Ath}{{A_{\ttbar}^\mrm{high} }}
\newcommand{\Atl}{{A_{\ttbar}^\mrm{low} }}
\newcommand{\ptl}{{p^{l}_{T}}}
\newcommand{\Atcum}{A_{\ttbar}\,(p^l_{T,\mrm{cut}})}
\newcommand{\Alcum}{A_{l}\,(p^l_{T,\mrm{cut}})}
\newcommand{\fbinv}{\mrm{\:fb}^{-1}}

\title{\boldmath\sffamily\bfseries\Large%
  Data driving the top quark forward--backward asymmetry with a
  lepton-based handle
  \unboldmath}
\author{\hphantom{MisterX}\\[-4mm]\sffamily\bfseries\normalsize%
  Adam~Falkowski,$^1$\thanks{adam.falkowski@th.u-psud.fr}\;\
  Michelangelo~L.~Mangano,$^2$\thanks{michelangelo.mangano@cern.ch}\;\
  Adam Martin,$^{2,3}$\thanks{adam.martin@cern.ch}
  \\[2mm]\sffamily\bfseries\normalsize%
  Gilad Perez$\hphantom{,}^{2,4}$\thanks{gilad.perez@cern.ch}\;\,\
  and\;\ Jan Winter$\hphantom{,}^{5,2}$\thanks{jan.winter@cern.ch}\\[4mm]
  {\small\sl$^1\,$Laboratoire de Physique Theorique d'Orsay, UMR8627-CNRS,
    Universite Paris-Sud, Orsay, France}\\[-2pt]
  {\small\sl$^2\,$PH-TH Department, CERN, CH-1211 Geneva 23, Switzerland}\\[-2pt]
  {\small\sl$^3\,$Department of Physics, University of Notre Dame,
    Notre Dame, IN 46556, USA\,}~\thanks{visiting scholar}\\[-2pt]
  {\small\sl$^4\,$Department of Particle Physics and Astrophysics,
    Weizmann Institute of Science, Rehovot 76100, Israel}\\[-2pt]
  {\small\sl$^5\,$Max-Planck-Institute for Physics, F\"ohringer Ring 6, 
    D-80805 Munich, Germany}
}
\date{\vskip4pt{\sffamily\bfseries December, 2012~}\\\small(\today~version)}

\begin{document}
\renewcommand{\baselinestretch}{1.2}
\maketitle
\renewcommand{\baselinestretch}{1.1}
\thispagestyle{empty}\vspace*{-8mm}

\begin{abstract}
\noindent
We propose that, within the standard model, the correlation between the
$\ttbar$\/ forward--backward asymmetry $A_{t\bar t}$ and the
corresponding lepton-based asymmetry $A_l$ -- at the differential
level -- is strong and rather clean both theoretically and
experimentally. Hence a combined measurement of the two distributions
as a function of the lepton $p_T$, a direct and
experimentally clean observable, would lead to a potentially
unbiased and normalization-free test of the standard model prediction.
To check the robustness of our proposal we study how the correlation
is affected by mis-measurement of the $t\bar t$\/ system transverse
momenta, acceptance cuts, scale dependence and compare the results of
\mcfm, \powheg (with \& without \pythia showering), and \sherpa's
\cssr in first-emission mode. We find that the shape of the relative
differential distribution $A_{l}\,(p^{l}_{T})\,[A_{\ttbar}\,(p^l_T)]$
is only moderately distorted hence supporting the usefulness of our
proposal. Beyond the first emission, we find that the correlation is
not accurately captured by lowest-order treatment. We also briefly
consider other differential variables such as the system transverse
mass and the canonical $t\bar t$\/ invariant mass. Finally, we study
new physics scenarios where the correlation is significantly distorted
and therefore can be more readily constrained or discovered using our
method.
\end{abstract}

\begin{flushright}
  \vspace*{-211mm}{\small CERN-PH-TH/2012-355\\LPT~12-114\\MPP-2012-183}
\end{flushright}

\clearpage%\newpage
%\renewcommand{\baselinestretch}{0.7}
%\tableofcontents
%\thispagestyle{empty}
%\clearpage
\renewcommand{\baselinestretch}{1.04}
%\vspace*{4mm}
%\noindent\hrulefill
%\vspace*{7mm}

% ======= intro ===============================================================

\section{Introduction}\label{sec:intro}
%%%%%%%%%%%%%%%%%%%%%%%%%%%%%%%%%%%%%%%%%%%%%%%%%%% 
Within the
standard model (SM), the $\ttbar$\/ forward--backward asymmetry, $\At$,
is an interesting variable because it tells us about QCD interactions
beyond leading order but in a region that should be well described by
perturbation theory~\cite{Kuhn:1998jr,Kuhn:1998kw}. Furthermore, as
the standard model contributions are expected to be
small~\cite{Kuhn:1998jr,Kuhn:1998kw,Bowen:2005ap,Antunano:2007da,Almeida:2008ug},
the measurement of $\At$ is sensitive to beyond-the-SM (BSM)
contributions. The asymmetry is quite an unique observable since
shifting it requires new physics with non-standard couplings both to
the $t\bar t$\/ quark current as well as to the to the current of
$u\bar u$\/ (or possibly $d\bar d\,$) initial-state
quarks.\footnote{Flavor-violating, $t$-channel new physics mechanisms
  to shift $\At$ can be Fierz rearranged into a form where this is true.}

The current status of top quark asymmetry related measurements at the
Tevatron is intriguing. It is useful to classify the current
data into measurements that directly probe the $t\bar t$\/ asymmetries
and measurements that probe daughter asymmetries, such as the
lepton-based ones. The asymmetries are quoted at several different
stages of the analysis. The easiest number to compare with theory is
the ``unfolded'', or ``production level'' asymmetry, where the
collaborations have processed the measured asymmetry to remove
background contamination, the effects of analysis cuts and of the
detector. Both CDF~\cite{Aaltonen:2012it,Aaltonen:2011kc}
($9.4\fbinv$) and D\O~\cite{Abazov:2011rq} ($5.4\fbinv$) present
their inclusive semileptonic $\ttbar$\/ asymmetry results at this
level. The average of the two measurements is 
\beq
\At\;=\;0.174\pm 0.038\,,
\label{eq:expav}
\eeq
which is significantly larger than the SM prediction,
\beq
A^\mrm{SM}_{t\bar t}\;=\;0.088\pm 0.006\,,
\label{AFBsm}
\eeq
obtained from next-to-leading order (NLO) QCD and including the
leading electroweak (EW) contributions~\cite{Kuhn:1998jr,Kuhn:1998kw,Hollik:2011ps,Manohar:2012rs,Bernreuther:2012sx}. The SM prediction
is derived taking the leading order total cross section in the
denominator of the asymmetry -- a conservative approach -- and the
error on the SM prediction has been estimated by varying
renormalization and factorization scales (see \eg Ref.~\cite{Drobnak:2012rb}).

Both CDF and D\O\ have also measured the dependence of the asymmetry
on the mass and rapidity of the $\ttbar$\/ system. For the $m_{\ttbar}$
dependence, CDF~\cite{Aaltonen:2012it, Aaltonen:2011kc,AFBCDF1} finds,
after unfolding:
\begin{align}
\Atl &\;\equiv\;\At\,(m_{\ttbar} < 450\gev)\;=\;0.084 \pm 0.053\,,\nn\\
\Ath &\;\equiv\;\At\,(m_{\ttbar} > 450\gev)\;=\;0.295 \pm 0.066\,.
\label{AtthCDF}
\end{align}
The asymmetry in the high-$m_{\ttbar}$ bin is particularly striking
given that the SM prediction, including EW
corrections~\cite{Kuhn:1998jr,Kuhn:1998kw,Hollik:2011ps,Manohar:2012rs,Bernreuther:2012sx},
is much lower:
\beq 
(\Ath)^{\rm SM}\;\equiv\;0.129^{+0.008}_{-0.006}\,.
\label{AFBsmhigh}
\eeq
For the same quantity, D\O\ has only reported a measurement without
unfolding (``reconstruction level''):
\begin{align}
\left(\Ath\right)_{\rm reco}\;=\;0.115 \pm 0.060\,.
\end{align}
We note that the D\O\ value for $\Ath$ is consistent with the CDF
value at the reconstruction level, $(\Ath)_{\rm reco}=0.198\pm0.043$,
which suggests that upon unfolding the value obtained by D\O\ would be
larger than the SM expectation. Putting it differently, assuming the
same unfolding factor between CDF and D\O, we find that both
measurements prefer a rather large value for $\Ath$, but with
D\O\ being more consistent with the SM prediction. 

The other class of forward--backward asymmetric observables is the
lepton-based asymmetries. From the same selected events used to measure
$\At$, both Tevatron experiments have also measured the single, $\Al$,
and dilepton, $\All$, asymmetries~\cite{AFBCDF1,Abazov:2011rq,AFBCDF3,:2012bfa}.
The results, given at the unfolded level,\footnote{Note that the CDF
  collaboration has not yet published an unfolded result for $\Al$,
  the one we state here is given for events at the
  ``background-subtracted level''.}
are:
%\begin{subequations}
\begin{align}
\left(\Al\right)_\text{CDF}&\;=\;0.066 \pm 0.025\,,&
\left(\All\right)_\text{CDF}&\;=\;0.42 \pm 0.15\pm 0.050\,,&\\
\left(\Al\right)_\text{D\O}&\;=\;0.152 \pm 0.04\,,&
\left(\All\right)_\text{D\O}&\;=\;0.053 \pm 0.084\,.&
\end{align}
%\end{subequations}
The SM predictions for the leptonic asymmetries, as reported by the
experimental collaborations, are:
\begin{align}
\left(\Al\right)^\mrm{SM}_\text{CDF}&\;=\;0.016\,,&
\left(\All\right)^\mrm{SM}_\text{CDF}&\;=\;0.060 \pm 0.01\,,
\hphantom{\pm 0.050\!\!}&\\
\left(\Al\right)^\mrm{SM}_\text{D\O}&\;=\;0.021 \pm 0.001\,,&
\left(\All\right)^\mrm{SM}_\text{D\O}&\;=\;0.047 \pm 0.001\,.&
\end{align}

Being a proton--proton collider, the LHC is not sensitive to $\At$.
However, the LHC can probe a related observable -- the charge
asymmetry in $t\bar t$\/ production, $A_C$. Measurements of
$A_C$ at $\sqrt{s}=7\tev$ have been reported by both
ATLAS~\cite{ATLAS-CONF-2012-057} (for $4.7\fbinv$)  and
CMS~\cite{:2012xv,Chatrchyan:2011hk} (for $5.0\fbinv$).
The expected SM asymmetry~\cite{Kuhn:1998jr,Kuhn:1998kw,
  Bernreuther:2012sx}, $A_C^{\rm SM} = 0.0115 \pm 0.0006$, 
is much smaller than the Tevatron's asymmetries, due to the domination
of the gluon--gluon production channel, which is symmetric. The
LHC measurements so far are consistent with the SM value, but 
the size of the statistical and systematic uncertainties is such that
one cannot yet exclude the consistency with the anomalous Tevatron
$\At$ result. It is important to emphasize that even within the SM the
Tevatron and LHC observables differ in nature. In particular, the
dominant $\ttbar$\/ production mechanism and the kinematical reaches
available to the top quarks are clearly very different at the two
colliders; the Tevatron collides charge-asymmetric beams and top quark
production is dominated by quark--antiquark annihilation, while, at the
LHC, collisions are charge symmetric and top pair production is driven
by gluon--gluon collisions. Furthermore, non-SM dynamics can naturally
induce a large deviation for the forward--backward asymmetry at the
Tevatron without affecting the charge asymmetry at the
LHC~\cite{Drobnak:2012rb,Drobnak:2012cz,AguilarSaavedra:2011ci}.

Given that the LHC probes a different observable, we turn our
attention back to the Tevatron. The discrepancy between the SM
predictions and the measured asymmetries at the Tevatron could be due
to an unknown QCD effect, or an unidentified experimental
bias. Alternatively, it might be a hint of dynamics beyond the SM (for
a review, see \eg Ref.~\cite{Kamenik:2011wt}). Either way, the current
situation is not satisfying, and the main goal of this paper is to
investigate what other information can be used to gain more insight.
Specifically, we propose a correlation between $\At$ and daughter
asymmetries that are experimentally easy to measure and also under
theoretical control. This correlation can then be used to distinguish
the SM from more exotic explanations of $\At$.

Our basic idea is simple, at least in principle. In the SM, the
lepton-based asymmetry in $\ttbar$\/ events is completely determined by
the $\ttbar$\/ asymmetry, meaning for a given $\At$ one can use top
quark decay kinematics to predict $\Al$. Radiation originating from the
top quark decay products alters the kinematics and blurs the
relationship between $\Al$ and $\At$, however this effect is
suppressed by the narrow width of the top quark. This relationship is
true for the inclusive asymmetries, but also differentially -- taking
the asymmetries with respect to a kinematic variable $x$; in each bin
of $x$, the lepton asymmetry can be fixed knowing $\At$ in that bin,
such that $\Al\,(x)[\At\,(x)]$ traces a calculable curve as $x$\/ is
varied.

However, once we move beyond the SM, $\Al$ and $\At$ are generically
independent. At high $m_{\ttbar}$, $\Al$ is indeed driven by the top
quark kinematics and polarization~\cite{Agashe:2006hk,Almeida:2008tp,Choudhury:2010cd, Krohn:2011tw,Berger:2012nw,Berger:2012tj,Cao:2010nw}, however,
near the $\ttbar$\/ threshold $\Al$ is set by the initial-state quark
polarization rather than anything related to the top
quarks~\cite{Falkowski:2011zr}. Thus, given some observable $x$\/ that
interpolates between the threshold and high-$m_{\ttbar}$ regions
(lepton $p_T$, $H_T$, etc.), the curve $\Al(x)$ versus $\At(x)$ will be
different for models beyond the SM. Our proposal is to use $x=p^l_T$
(the lepton $p_T$) and to simultaneously measure $\Alp$ and $\Atp$, to
verify whether the curve $\Alp[\Atp]$ is in agreement with the SM. We
choose the lepton $p_T$ as our kinematic variable because it is
experimentally clean and easy to reconstruct.

We begin our study of this correlation in Sec.~\ref{sec:ideal},
working at the parton level and without cuts to demonstrate the basic
idea. In Sec.~\ref{sec:pleveltests} we provide several checks that
suggest that the correlation is indeed robust and therefore more
sensitive to new physics contributions to the asymmetries. In detail,
we consider three types of effects: (1) mis-modeling of the of the
$\ttbar$\/ transverse momenta, (2) scale dependence of the
differential asymmetries and (3) radiation in the top decay. We verify
that, while some of these effects influence the overall normalization
of the asymmetries, affecting the agreement between theory and
experiment, the correlation between $\Alp$ and $\Atp$ is unaffected by
these deformations. Then, in Sec.~\ref{sec:realistic} we include
detector acceptance, experimental cuts, parton showering, and top
quark reconstruction, to show the asymmetry correlation in a realistic
hadron-collider environment. Finally, in Sec.~\ref{sec:BSM} we
consider several simple new physics models and show that the SM
correlation is significantly violated in general which therefore can
potentially lead to much cleaner extraction of a possible non-SM
signal. This is followed by some discussion regarding the use of
reconstruction-free variables (Sec.~\ref{sec:discussion}) and our
conclusions (Sec.~\ref{sec:conclusions}).

\section{Idealized case: SM}\label{sec:ideal}

To get some intuition regarding our proposal we begin by discussing
the idealized SM case where no acceptance cuts are included.
The differential asymmetry observables are defined as:
\begin{align}
\label{eq:diff}
\Atp&\;=\;\frac{N_{\upDelta Y_{\ttbar}>0}\,(\ptl) - N_{\upDelta Y_{\ttbar}<0}\,(\ptl)}
{N_{\upDelta Y_{\ttbar}>0}\,(\ptl) + N_{\upDelta Y_{\ttbar}<0}\,(\ptl)}\,,\\[5mm]
\Alp&\;=\;\frac{N_{Y_l>0}\,(\ptl) - N_{Y_l<0}\,(\ptl)}
{N_{Y_l>0}\,(\ptl) + N_{Y_l<0}\,(\ptl)}\,,
\end{align}
where $N_{\upDelta Y_{\ttbar}>0}$ ($N_{Y_l>0}$) and $N_{\upDelta Y_{\ttbar}<0}$
($N_{Y_l<0}$) are the number of events with $\upDelta Y_{\ttbar}$
($Y_l$) greater or less than zero.\footnote{$Y_l$ is defined as
  $Q_l\cdot\eta_l$, such that a backwards-moving electron is the
  same as a forwards-moving positron.}
We also study the {\em cumulative} distributions -- the asymmetry for
all events with lepton $p_T$ above a given threshold, obtainable by
integrating the numerator and denominator of the differential
distributions, then taking the ratio:
\begin{align}
\Atcum&\;=\;\dfrac{\displaystyle\int\limits_{p_{T,\mrm{cut}}}^{\infty}\left(
  N_{\upDelta Y_{\ttbar}>0}\,(\ptl) - N_{\upDelta Y_{\ttbar}<0}\,(\ptl)\right)}{
  \displaystyle\int\limits_{p_{T,\mrm{cut}}}^{\infty}\left(
  N_{\upDelta Y_{\ttbar}>0}\,(\ptl) + N_{\upDelta Y_{\ttbar}<0}\,(\ptl)\right)}\,,\\[4mm]
\Alcum&\;=\;\dfrac{\displaystyle\int\limits_{p_{T,\mrm{cut}}}^{\infty}\left(
  N_{Y_l>0}\,(\ptl) - N_{Y_l<0}\,(\ptl)\right)}{
  \displaystyle\int\limits_{p_{T,\mrm{cut}}}^{\infty}\left(
  N_{Y_l>0}\,(\ptl) + N_{Y_l<0}\,(\ptl)\right)}\,.
\label{eq:cum}
\end{align}
The differential distributions contain the physics we want to study --
the correlation between $\Al$ and $\At$, but they are difficult to
measure experimentally owing to the limited top quark sample size.
Meanwhile, cumulative distributions are more tractable experimentally,
but the integration over multiple bins dilutes the correlation between
$\At$ and $\Al$. We present both types of distributions for the
idealized SM case to show the similarities and differences.

We are interested in the lepton asymmetry in the lab frame, as well as
the lepton asymmetry after boosting to a frame where the $\ttbar$\/
system has no longitudinal momentum. The lepton kinematics, which
encode the asymmetry inherited from the top quarks, get smeared under
motion of the $\ttbar$\/ system, hence boosting back leads to a larger
$\Al$. The boost only effects the leptonic asymmetry, as $\At$ is
defined in terms of a rapidity difference and is manifestly invariant
under longitudinal boosts. Unless otherwise specified, we will use
the generic $\Al$ for the lab frame lepton asymmetry $A^\mrm{lab}_l$,
and use $A^\mrm{boost}_l$ to refer specifically to the lepton
asymmetry in the boosted frame.

The primary tools for our study are the NLO Monte Carlo generators
\mcfm(v6.3)~\cite{Campbell:2011bn,Campbell:2012uf}, and \powheg (here
run in the hardest-emission generator mode) using the heavy quark
production routines~\cite{Nason:2004rx,Frixione:2007vw,Alioli:2010xd,Frixione:2007nw}.
For the idealized SM case, all results were generated using the
MSTW2008NLO \cite{Martin:2009iq} parton distribution functions and
with factorization and renormalization scales set to
$\mu_\mrm{R}=\mu_\mrm{F}=Q=\sqrt{m^2_t+(p^t_T)^2}$. Spin
correlations between the top (antitop) quark and its corresponding
decay products are maintained in both codes.\footnote{Other relevant
  choices of Monte Carlo generation parameters are $m_t=173.0\gev$,
  $\mrm{\Gamma}_t=1.31\gev$ and $m_b=5.0\gev$.}

\begin{figure}[t!]
\centering\vskip4mm
\includegraphics[clip,width=0.44\textwidth]{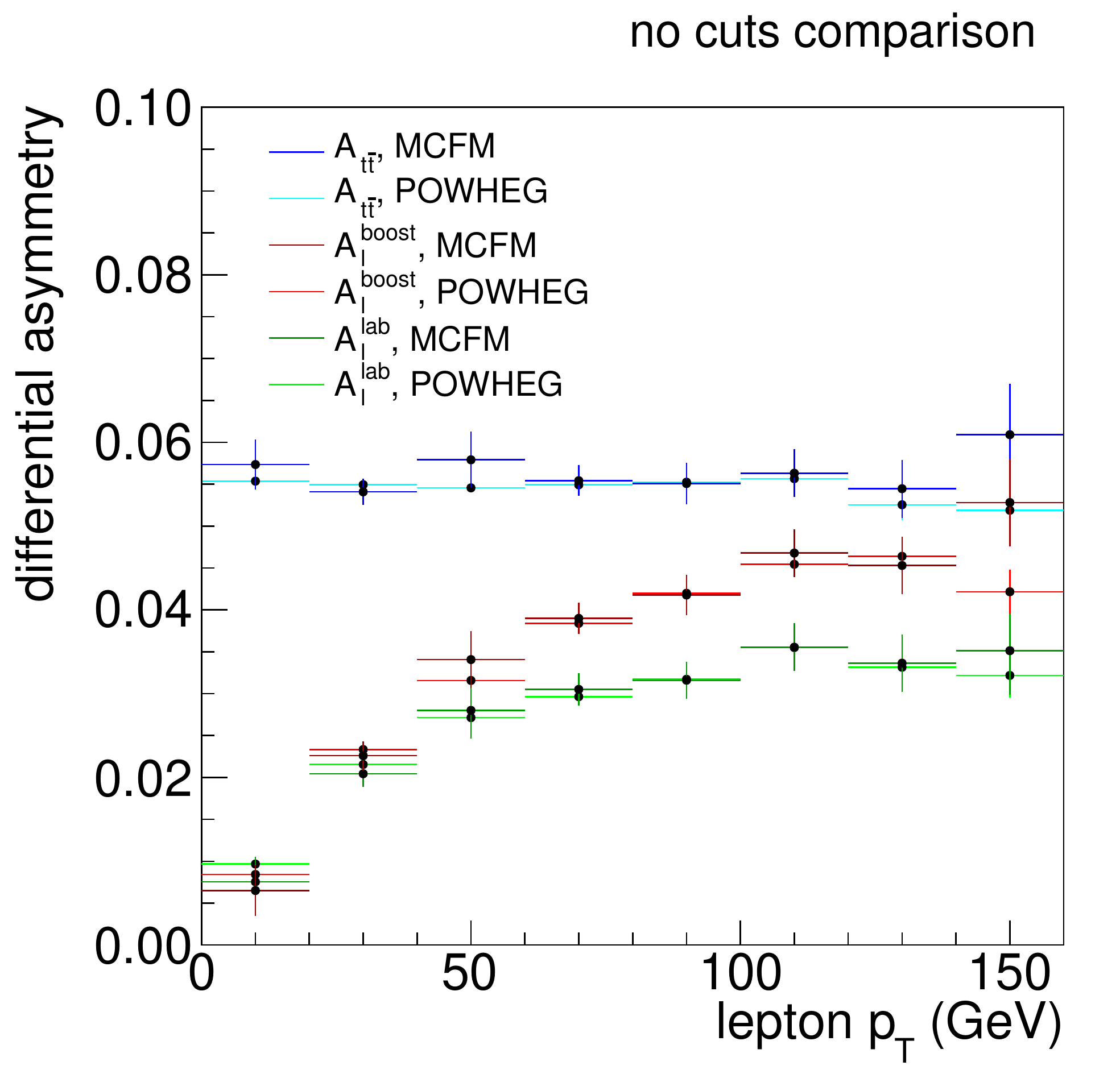}
\includegraphics[clip,width=0.44\textwidth]{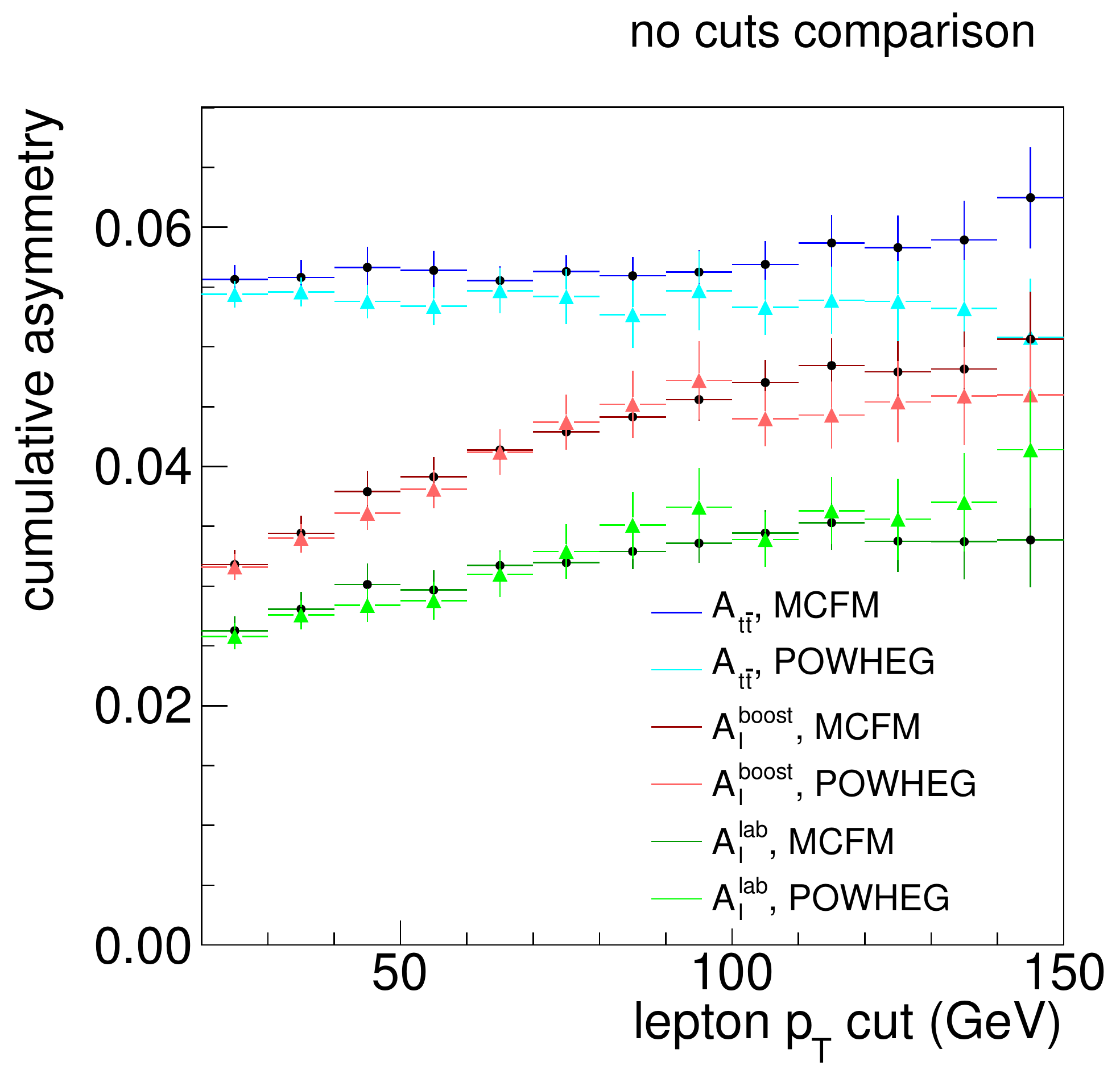} 
\caption{\label{fig:ideal}
  The dependence of top quark and leptonic asymmetries on the
  lepton $p_T$. The left panel shows the differential distributions
  $\Atp$, $\Alp$ for the SM ideal case, while the cumulative asymmetries
  $\Atcum$, $\Alcum$ are shown in the right hand panel. The darker lines
  depict the \mcfm results, while the results obtained with \powheg are
  shown in the lighter shaded lines. The green curves show $\Al$ in
  the lab frame, while the red curves show $\Al$ after boosting to a
  frame where the $\ttbar$\/ system has no longitudinal momentum. This
  boost only affects the leptonic asymmetries.}
\end{figure}

The distributions $\Atp$, $\Alp$, $\Atcum$ and $\Alcum$ for the ideal
case are shown below in Fig.~\ref{fig:ideal}. The darker lines show
the \mcfm results, while the results obtained with \powheg are shown in
the lighter shaded lines. The green curves show $\Alp$ in the lab
frame, while the red curves show $\Alpb$, the lepton asymmetry in the
$Y_{\ttbar}=0$ frame. We find that the two NLO Monte Carlo (MC)
generators are in reasonable agreement. The qualitative behavior of
the curves can be understood as follows: beginning with the
leptonic asymmetries, near threshold $\Alp$ is sensitive to the
polarization of the incoming quark, which is small due to the
vector-like nature of QCD. Hence we expect $\Al$ to be near
zero~\cite{Falkowski:2011zr}. In the other extreme limit, when the
lepton's $p_T$ is very large it has to come from a boosted top quark,
and therefore the lepton-based asymmetry should asymptote to the
corresponding value of $\At$, keeping in mind that within the SM no
net polarization is expected for the top quarks in $\ttbar$\/ events.
This is consistent with the lepton-based asymmetry curves shown in the plot.

\begin{figure}[b!]
\centering\vskip7mm
\includegraphics[clip,width=0.44\textwidth]{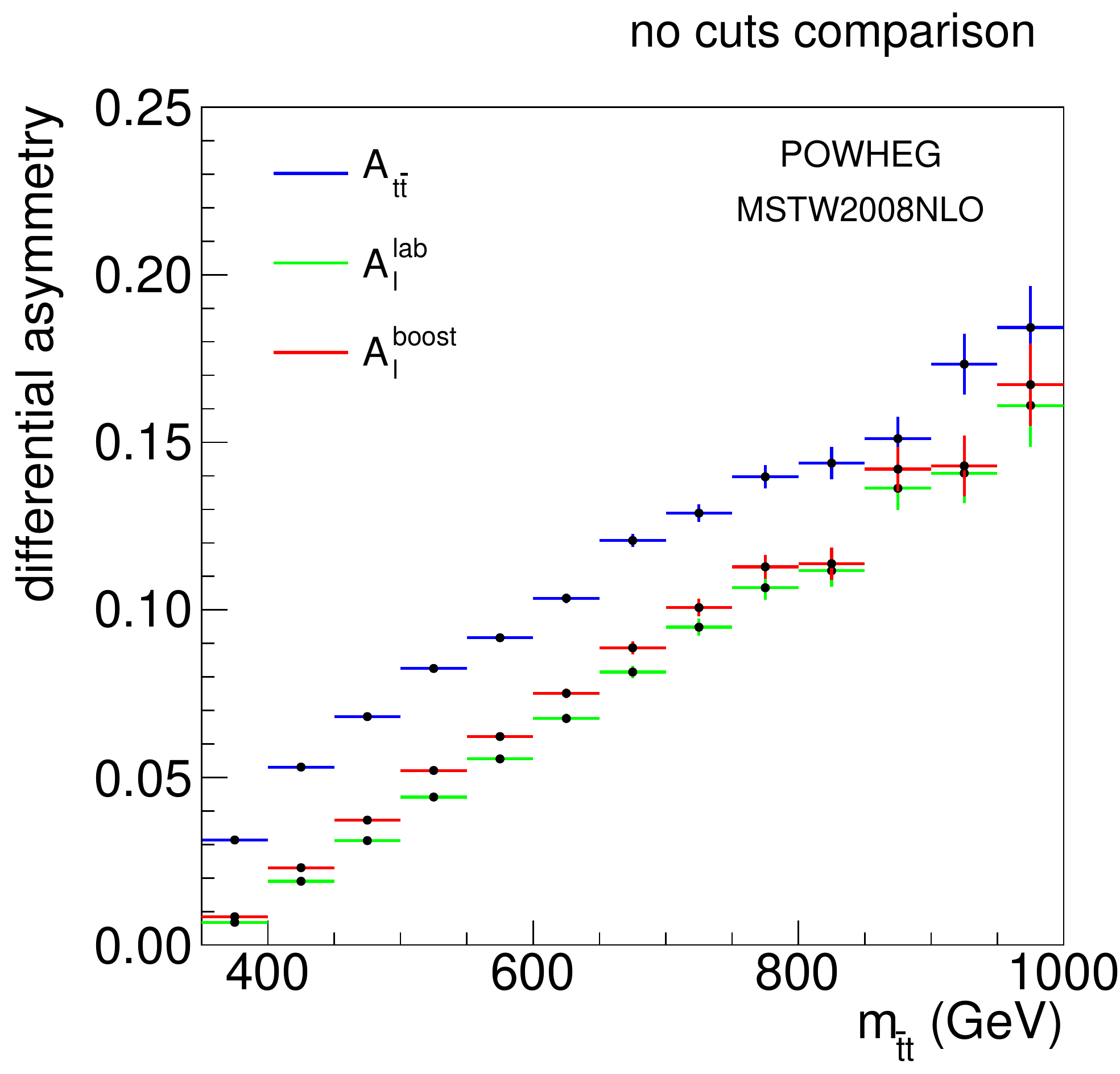}
\includegraphics[clip,width=0.44\textwidth]{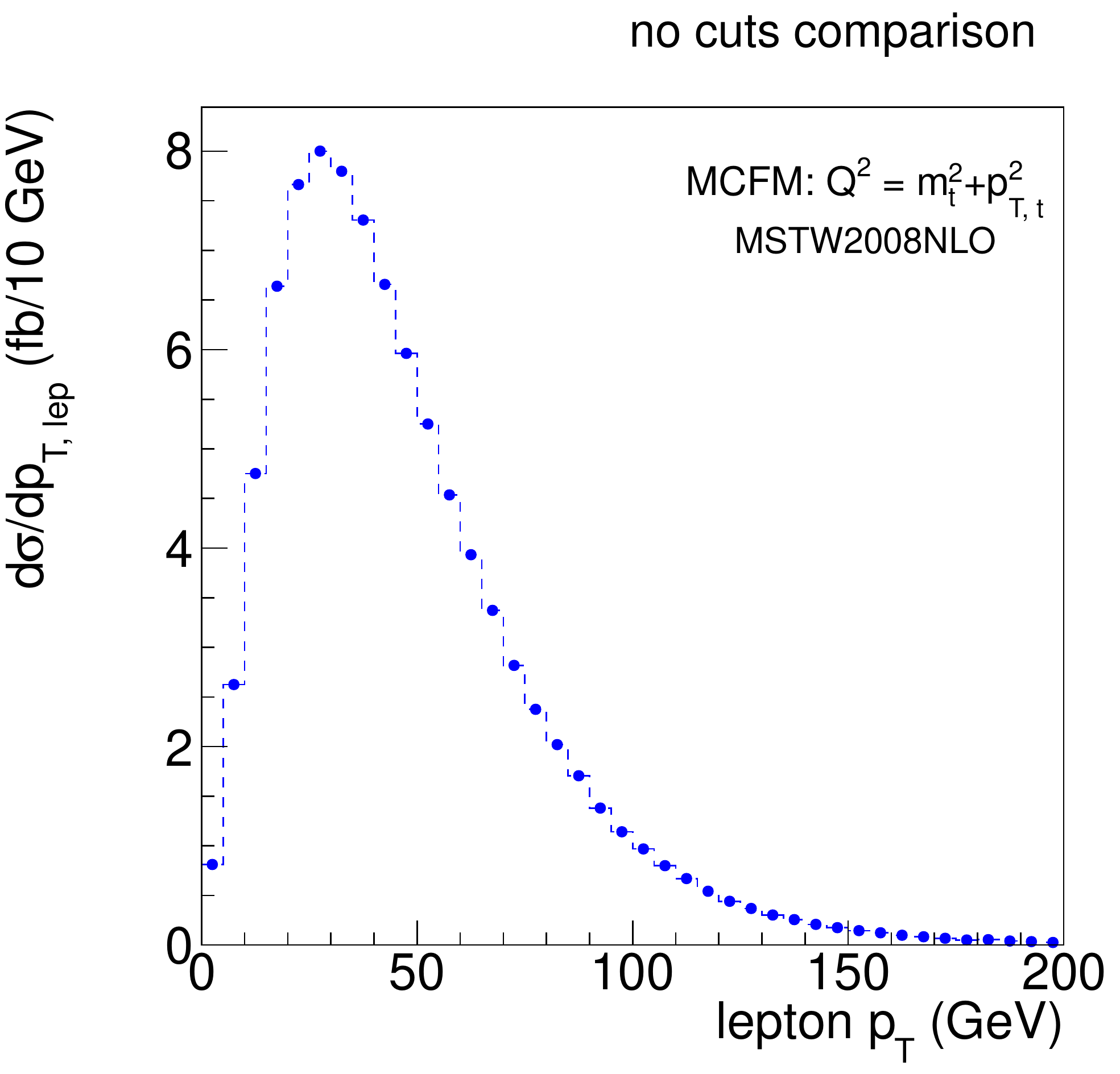}
\caption{\label{figmtt}
  The differential asymmetries $\Atm, \Alm$ for the SM ideal
  case are shown in the left panel. The lepton $p_T$ distribution (at
  LO) is shown in the right panel, normalized to the total cross
  section.}
\end{figure}

\begin{figure}[b!]
\centering\vskip2mm
\includegraphics[clip,width=0.44\textwidth,height=0.41\textwidth]{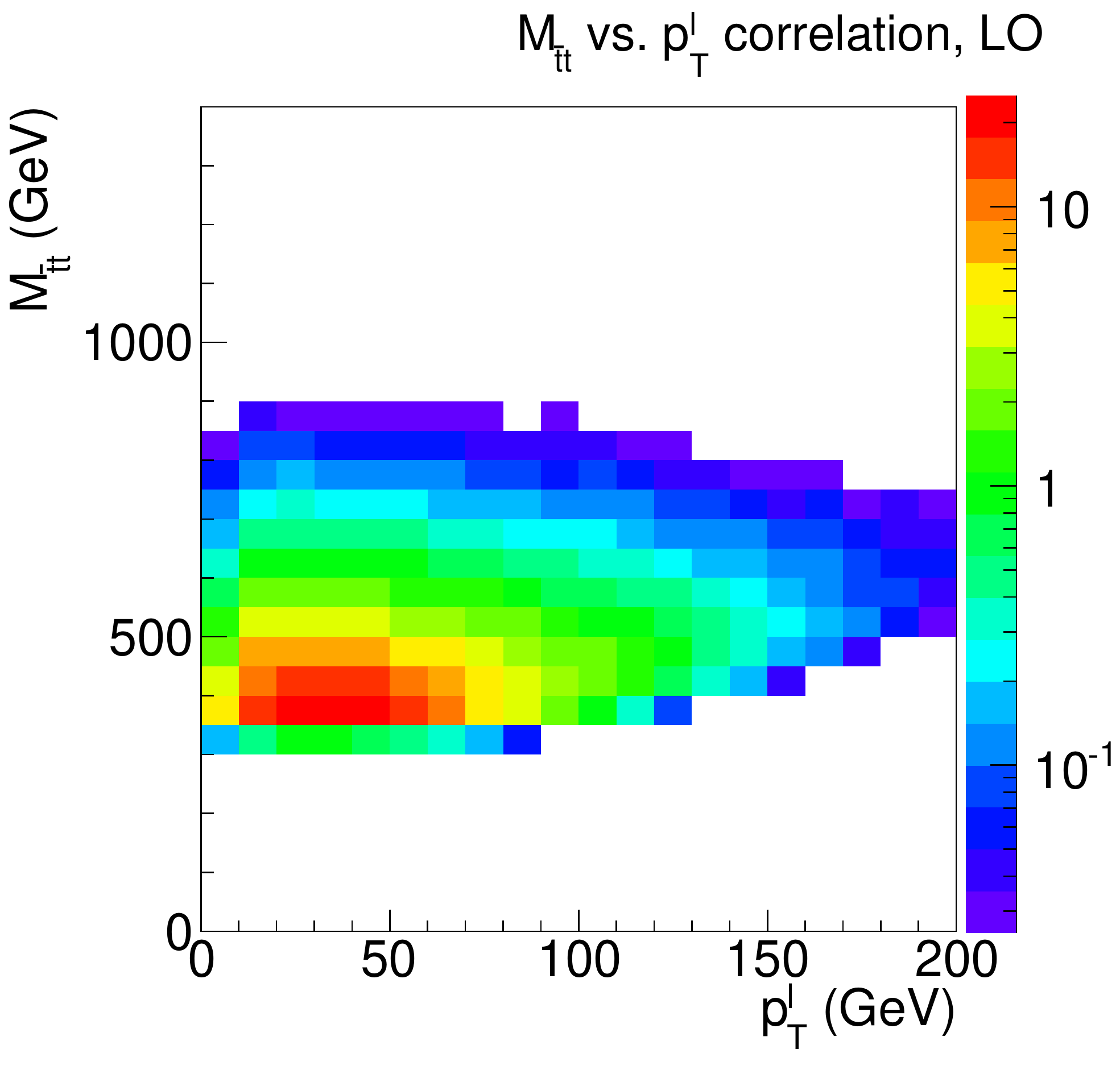}
\includegraphics[clip,width=0.44\textwidth,height=0.41\textwidth]{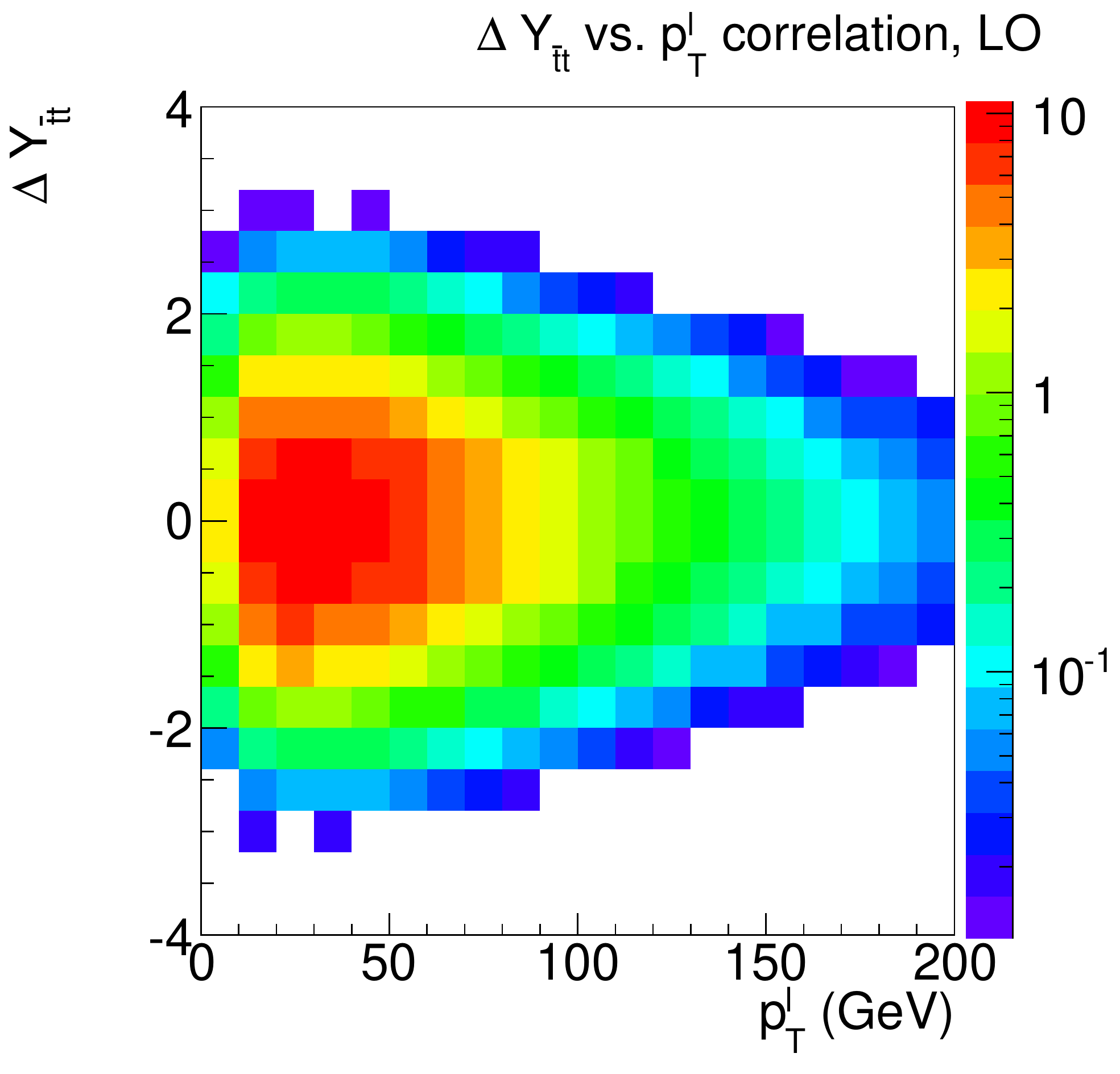}
\caption{\label{cor}
  The correlation between the lepton $p_T$ and the kinematics
  of the $\ttbar$\/ system (at LO). The correlation between $\ptl$ and
  $m_{\ttbar}$ is shown on the left, and the correlation between
  $\ptl$ and $\upDelta Y_{\ttbar}$ is shown on the right.}
\end{figure}

The behavior of $\Atp$ can be understood from the lepton $p_T$
spectrum in $\ttbar$\/ events and the dependence of $\At$ on $m_{\ttbar}$.
The asymmetry $\Atm$ is a monotonically increasing function of $\mtt$
(see Fig.~\ref{figmtt} where for completeness $\Atm$ and $\Alm$ are
presented)~\cite{Almeida:2008ug,Ahrens:2011uf,Kidonakis:2011ca},
however $\mtt$ is only weakly correlated with $\ptl$ (at least up to
$\ptl\lesssim100\gev$). The correlation, at leading order (LO),
between these variables is shown explicitly in Fig.~\ref{cor}. The
weakness of the $m_{\ttbar}-\ptl$ correlation implies that the lower
$\ptl$ bins are populated almost equally by a wide range of invariant
masses and hence we do not expect a significant rise in $\Atp$ as the
lepton $p_T$ is varied.\footnote{A similar argument for the flatness of $\At\,(\ptl)$ can be made using the $\upDelta Y_{\ttbar}-\ptl$ correlation (at LO) depicted in the right panel of Fig.~\ref{cor}. The differential asymmetry $\At\,(|\upDelta Y_{\ttbar}|)$ rises monotonically/linearly
  for increasing absolute values of $\upDelta Y_{\ttbar}$, and from Fig.~\ref{cor} we see that each $\ptl$ bin picks events with a variety of $\upDelta Y_{\ttbar}$ values. The asymmetry in a given $\ptl$ bin is thus the average of small $\At$ at low $|\upDelta Y_{\ttbar}|$
  with large $\At$ at high $|\upDelta Y_{\ttbar}|$. As $\ptl$ is increased, the sampling across a range of $\upDelta Y_{\ttbar}$ stays, but the cross section decreases. However, as the change in cross section cancels out in the ratio defining the asymmetry (a normalization effect), $\At$ remains flat.}
 
Higher $\ptl$ events are more correlated with large $\mtt$ however
the large lepton transverse momenta, which naively would lead to a large
$\At$, are forcing those energetic events to be central. As shown in
Fig.~\ref{cor} events with $\ptl\gtrsim 150\gev$ have $\mtt\gtrsim550\gev$,
however, with a top quark rapidity difference below $0.7$ (as shown in
Fig.~\ref{cor} visualizing the correlation between $\upDelta Y_{\ttbar}$
and $\ptl$). As central events tend to have lower $\At$, we actually expect the
overall value of $\At$ and $\Al$ to be below their nominal value
expected based on the inclusive and high invariant mass values prior
to lepton transverse momenta cuts (Fig.~\ref{figmtt}). This is
consistent with the distributions shown in Fig.~\ref{fig:ideal}.

\begin{figure}[t!]
\centering\vskip4mm
\includegraphics[clip,width=0.44\textwidth]{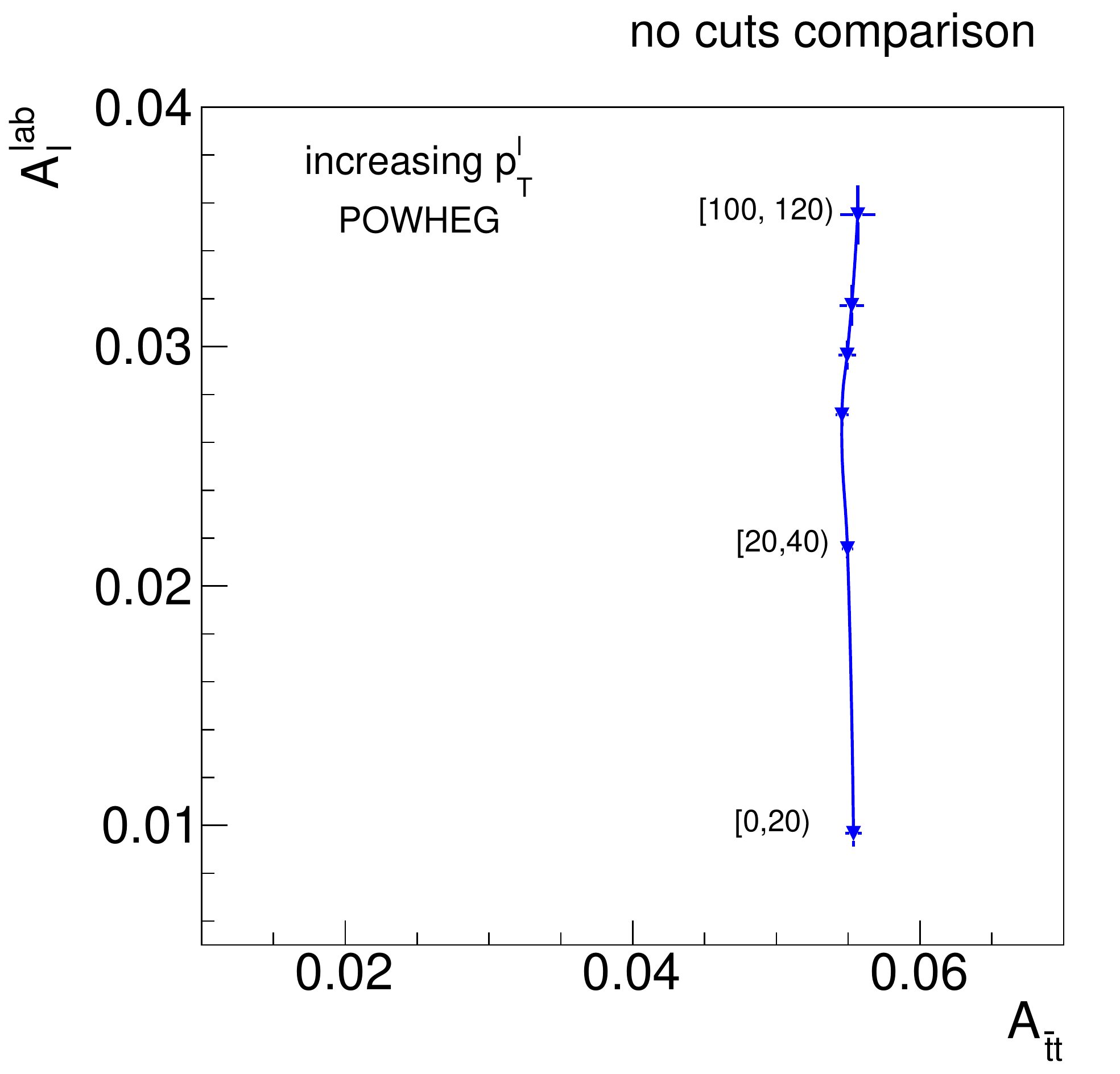}
\includegraphics[clip,width=0.44\textwidth]{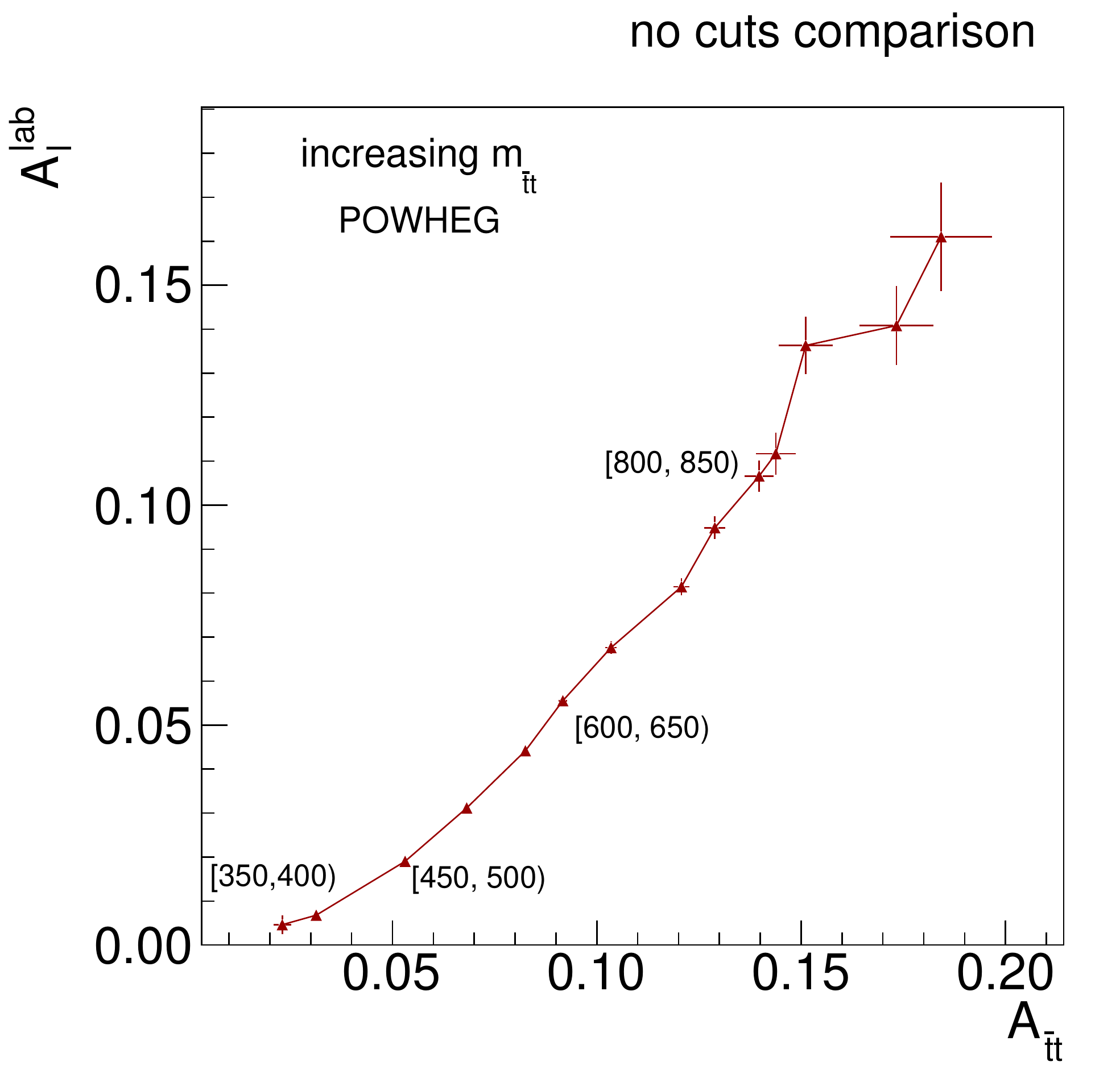}
\caption{\label{fig:attal_plane}
  The correlation between the leptonic and $\ttbar$\/ asymmetries
  displayed in the $\Al$ versus $\At$ plane. No cuts have been imposed,
  and all quantities have been calculated at parton level. In the left
  panel we show the curve traced out as $\ptl$ is increased
  from $0\gev$ (bottom point) to $100\gev$ (top point) in intervals of
  $20\gev$ -- the $\ptl$ intervals for some of the points are
  indicated in brackets on the plot. Similarly, in the right panel we
  present the curve traced out as $m_{\ttbar}$ is increased from the
  threshold value (bottom left) to $1\tev$ (top right) in intervals of
  $50\gev$.}
\end{figure}

We close this subsection by presenting the same result but this time
on the $\Al-\At$ plane where each point in the left panel of
Fig.~\ref{fig:attal_plane} corresponds to a different $\ptl$ bin,
which is nothing but the $\Alp[\Atp]$ curve mentioned in the
introduction. Given the one-dimensional asymmetry spectra explained
above, we find that the distribution of points is consistent
with a nearly vertical line, and both $\Atp$ and $\Alp$ span rather
small values. For completeness, we also show, in the right panel of
Fig.~\ref{fig:attal_plane}, the asymmetry correlation as a function of
$\mtt$, even though this is not the main focus of this work.

%%%%%%%%%%%%%%%%%%%%%%%%%%%%%%%%%%%%%%%%%%%%%%%%%%%%%%%%%%%%%%%%%%%%%%%%%%%%%
\subsection{Robustness tests}\label{sec:pleveltests}
%%%%%%%%%%%%%%%%%%%%%%%%%%%%%%%%%%%%%%%%%%%%%%%%%%%%%%%%%%%%%%%%%%%%%%%%%%%%%

Having understood the interplay between the SM $\At$ and $\Al$ in an
idealized scenario, we now study how various effects impact the
$\Alp-\Atp$ correlations. We continue to work at parton level in this
section and will not include cuts. Acceptance cuts and more realistic
jet description will be studied in later sections.

\begin{figure}[t!]
\centering\vskip4mm
\includegraphics[clip,width=0.44\textwidth]{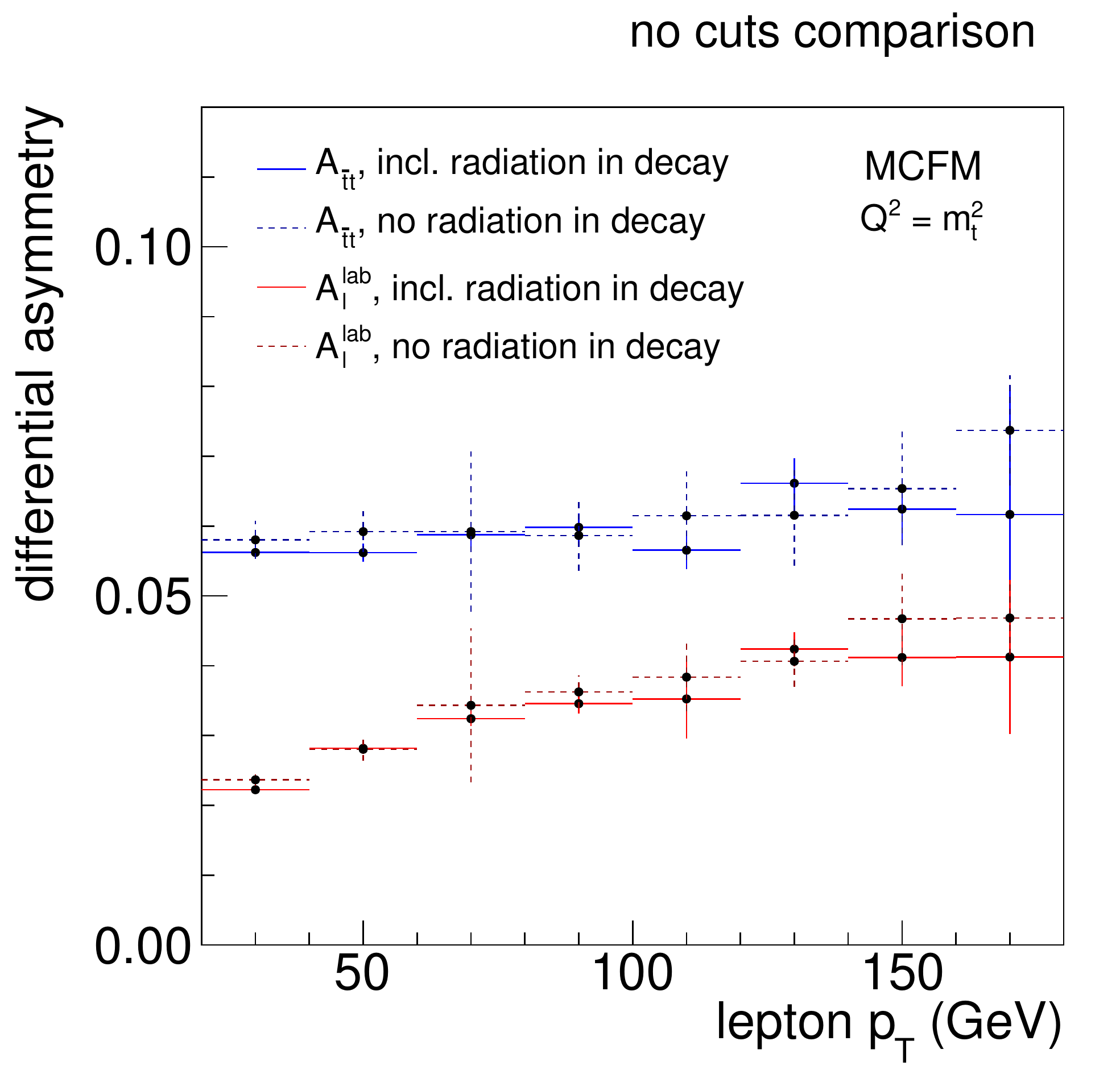}
\caption{\label{fig:mcfm_versioncompare}
  A comparison of \mcfm results with and without radiation in the top
  quark decay. The solid line shows the differential distributions
  $\Atp$ and $\Alp$ (lab frame) as a function of the lepton $p_T$
  in the most recent version (v6.3), while the dashed line shows the
  same observable calculated with the previous version. The most
  recent version includes NLO effects in decays, while the previous
  iteration contained all decay spin correlations, but no NLO effects
  in the decays.}
\end{figure}

The first effect we consider is radiation in decay. The $b$\/ ($\bar b$)
from the leptonic top (or antitop) quark can radiate gluons, changing the
kinematics and correlations among the top decay products.\footnote{The
  hadronic $W$'s decay products can also radiate, of course, but since
  the $W$\/ is both a color singlet and narrow, this has less influence
  on the lepton kinematics than the radiation from the lepton's
  sibling $b$\/ ($\bar b$) quark.}
Radiation in decay will obviously not change the top quark asymmetry,
but it may impact how $\At$ is passed on to $\Al$, \eg through the
analysis selection cuts. Though radiation in decay occurs at the same
order in $\alpha_\mrm{s}$ as the processes that contribute to $\At$, it is
further suppressed by the width of the top quark. Recent analytic work
on the effects of radiation in decay can be found in
Ref.~\cite{Melnikov:2011qx}, and radiation in decay has been
incorporated into the latest version of \mcfm \cite{Campbell:2012uf}.
By comparing the study of Fig.~\ref{fig:ideal} between different
versions of \mcfm, we can study the size of the effect of radiation in
decay on our observables $\Atp$ and $\Alp$. As can be seen from 
Fig.~\ref{fig:mcfm_versioncompare}, the results with and without
radiation in the decay are nearly identical, indicating that the LO
treatment of the top quark decay products (but with spin correlations
intact) is sufficient to predict the correlation between $\At$ and $\Al$.
As the top quark $p_T$ is ambiguous once radiation in decay is
allowed, we perform this cross-check using a fix-scale choice of
$Q^2=m^2_t$.

The second test of robustness concerns the $p_T$ of the $\ttbar$\/
system, $\pTt$. As is well known~\cite{Bowen:2005ap}, $\At$ strongly
depends on $\pTt$ as it controls the level of real emission in the
event. Therefore, among other effects, mis-modeling of acceptance cuts
or biases in the measurement could lead to a change in the overall
normalization of the resulting inclusive and differential value of
$\At$. Thus, it is important to verify whether the $\Al-\At$
correlation is sensitive to $\pTt$.
Notice that by insisting on large $\ptl$, we are forcing the
$\ttbar\,$+$X$\/ system into two possible kinematic configurations: (1)
the top and antitop quark move in the same direction, recoiling
against hard initial-state radiation, or (2) the top and antitop quark
are back-to-back and both are central. In the first case large lepton
$p_T$ are possible since some of the initial and large $\pTt$ is
inherited by the lepton. In the second configuration, the lepton
inherits large $p_T$ from the individual top or antitop quark, rather
than the $\ttbar$\/ pair. As a result, this configuration is
characterized by low $\pTt$, but large $m_{\ttbar}$. Given the
contributions come from such different kinematic regimes, one may
worry that our results in Fig.~\ref{fig:ideal} come from a delicate
cancellation between different effects. Any mis-modeling of or bias in
$\pTt$ would disrupt such a cancellation and destabilize the
correlation. One way to see whether a cancellation is occurring is to
divide the events into different bins of $\pTt$ and check the $\Alp-\Atp$
correlation in each bin. Again, computing $\ttbar$\/ production at NLO,
we have performed this test and the results are shown below in
Fig.~\ref{fig:ptcompare}. Quantifying the correlation by the ratio
$\Alp/\Atp$, for $p_{T,\ttbar}<20\gev$, we see it is the same as
the correlation in events with $p_{T,\ttbar}>20\gev$ -- a strong
indication that the $\Alp-\Atp$ correlation does not come from a
cancellation of competing effects. Based on our complete NLO analysis,
we can therefore conclude it is stable against $p_{T,\ttbar}$
mis-modeling.

\begin{figure}[t!]
\centering\vskip4mm
\includegraphics[clip,width=0.44\textwidth]{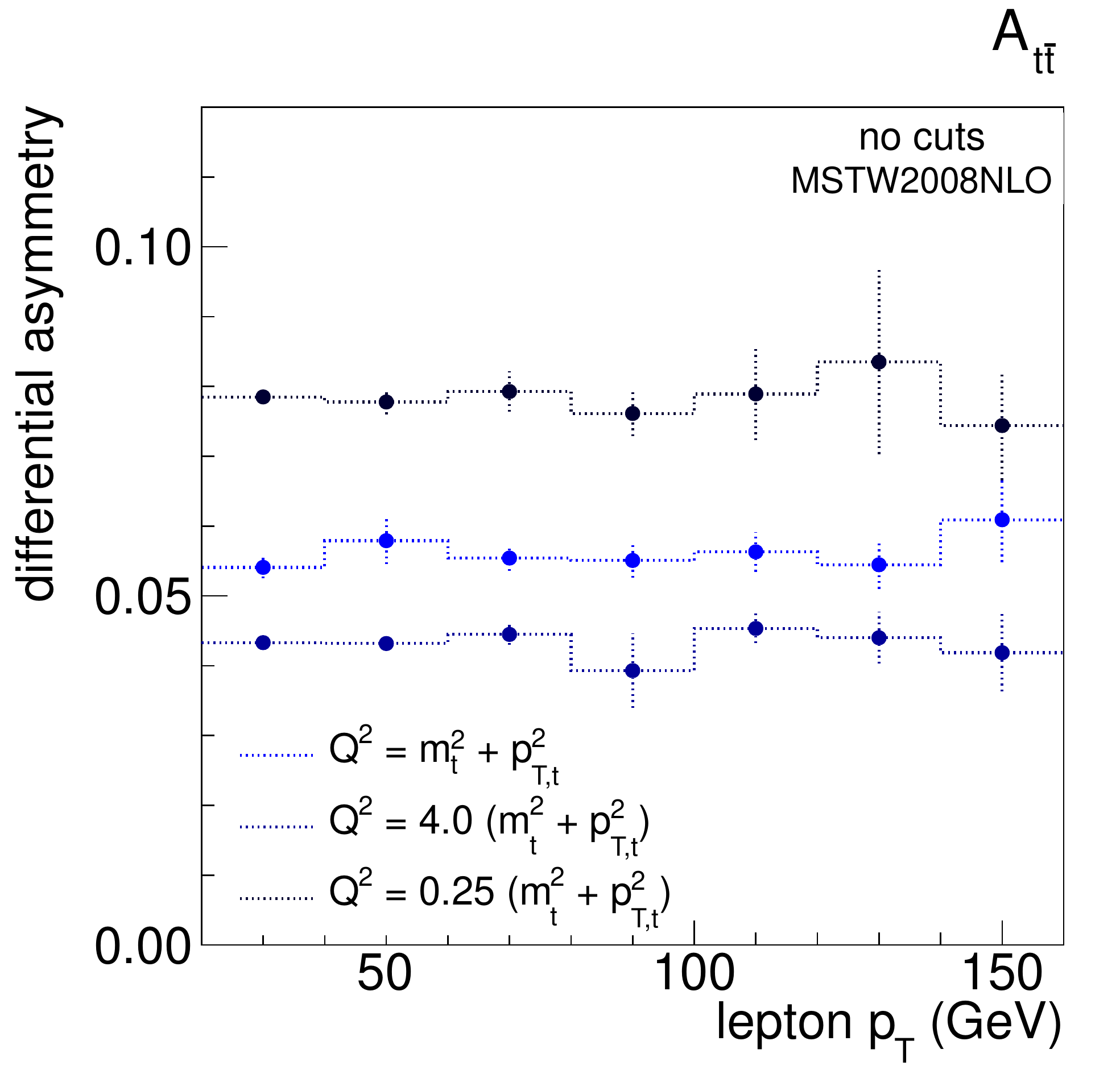}
\includegraphics[clip,width=0.44\textwidth]{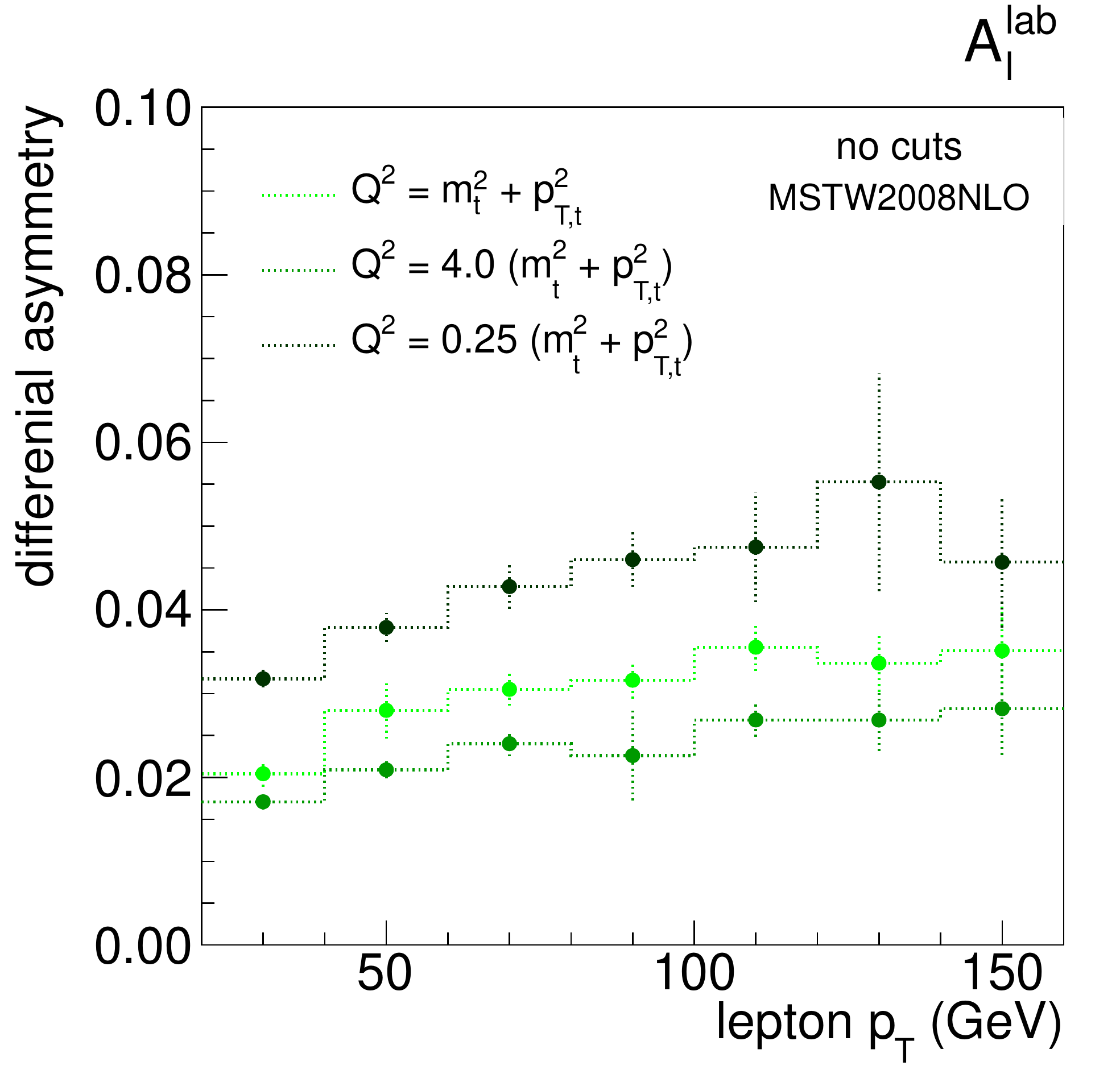}
\\\vskip2mm
\includegraphics[clip,width=0.44\textwidth]{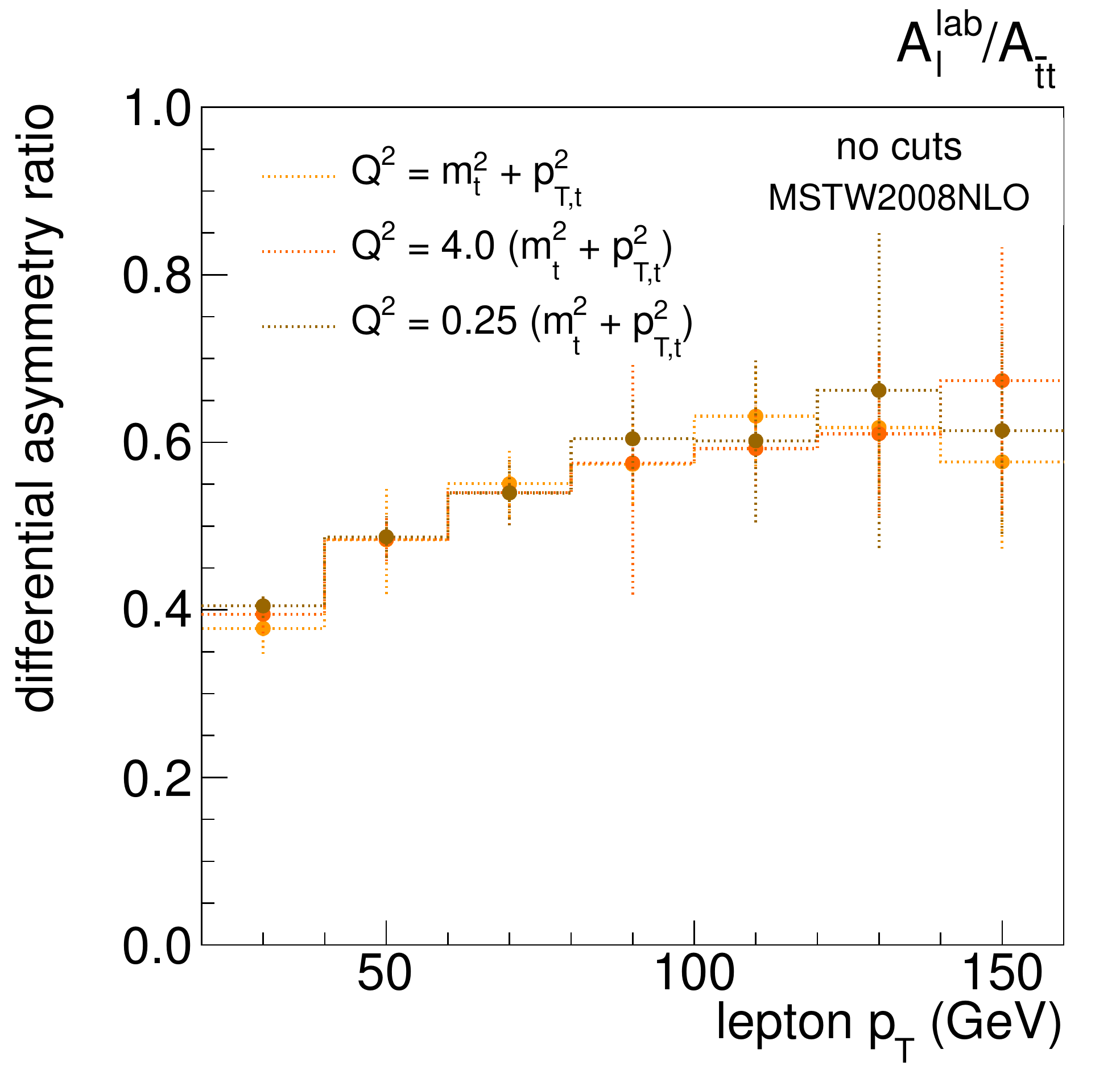}
\caption{\label{fig:scalevar}
  Dependence of the asymmetries on the lepton $p_T$ for three
  different scale choices, calculated using \mcfm. The left, middle
  and right panel show $\At$, $\Al$ (lab frame) and the ratio
  $\Al/\At$, respectively. Only the differential asymmetries are
  depicted, however the same trend is present at the cumulative level,
  cf.\ Eqs.~(\ref{eq:diff})--(\ref{eq:cum}). These plots show the
  ideal SM scenario where no cuts have been applied.}
\end{figure}

The final avenue we explore within the SM ideal case is scale
variation. Because $\At$ vanishes at ${\cal O}(\alpha^2_\mrm{s})$, NLO
calculations of differential $\ttbar$\/ properties only result in a
leading order prediction for the asymmetry $\At$. A more accurate
determination of the asymmetry would involve understanding $\ttbar$,
differentially, at NNLO. In the absence of this NNLO calculation (for
recent progress on this, see \eg
Refs.~\cite{Baernreuther:2012ws,Czakon:2012pz}), one estimate of our
ignorance regarding higher-order corrections is to vary the scale used
in the $\At$ calculation by a factor of two. While we expect the
absolute values of $\At$ and $\Al$ to change as the scale is varied,
much of the variation is carried in the scale where $\alpha_\mrm{s}$ is
evaluated, which cancels out if we take the ratio of asymmetries.  A
stable ratio $\Al/\At$ under scale variation is therefore a sign that
the correlation pointed out here is robust. In
Fig.~\ref{fig:scalevar} we show the differential distributions
$\Atp$, $\Alp$ and the ratio $\Alp/\Atp$ for three different scale
choices: $Q^2=Q^2_0$, $Q^2=4\times Q^2_0$ and $Q^2=Q^2_0/4$ where
$Q^2_0=m^2_t+(p^t_T)^2$. The ratio is indeed very stable, as can
clearly be seen from the bottom panel of Fig.~\ref{fig:scalevar}.

\begin{figure}[h!]
\centering\vskip7mm
\includegraphics[clip,width=0.44\textwidth]{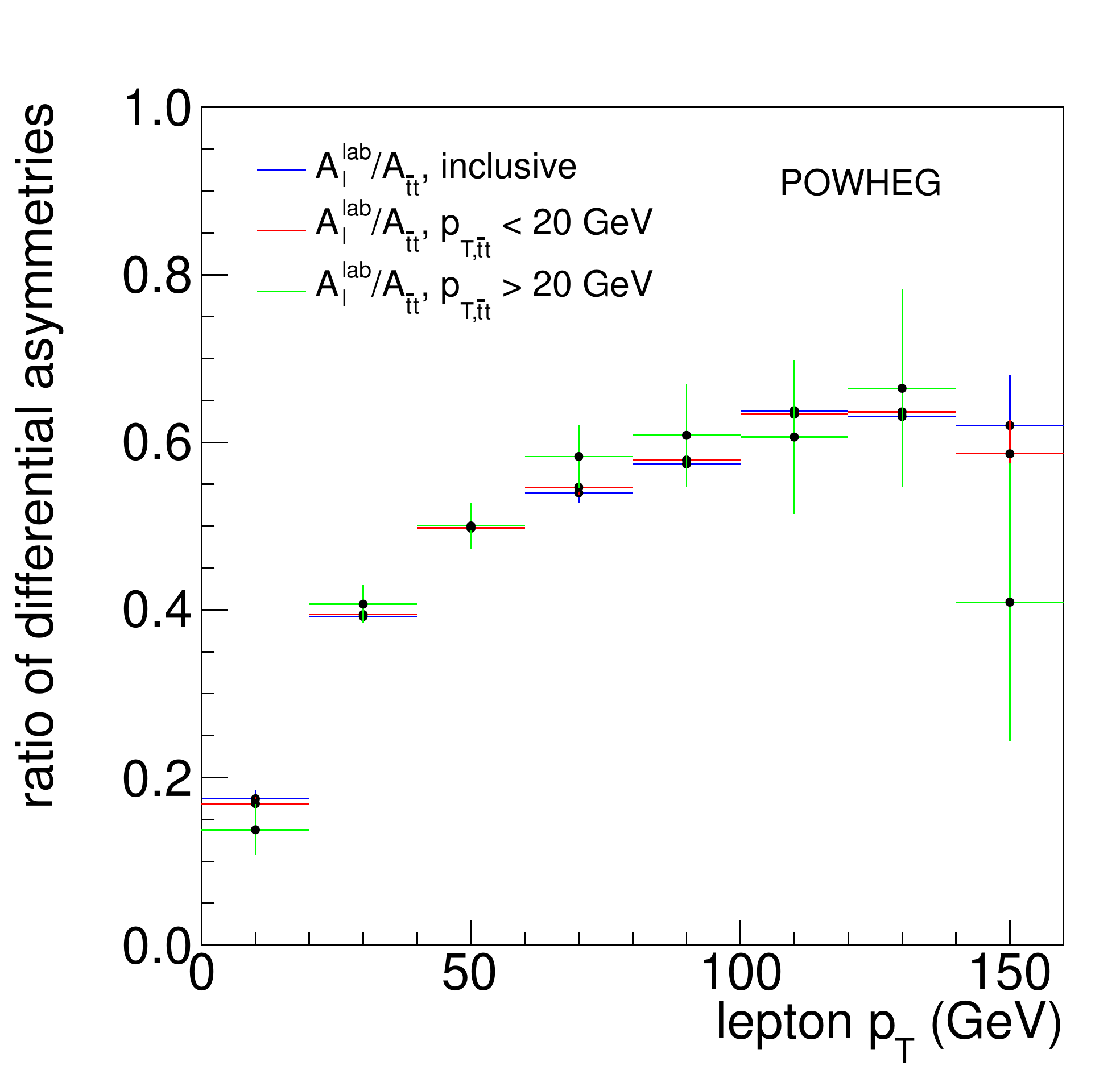} 
\caption{\label{fig:ptcompare}
  The ratio of differential asymmetries $\Alp/\Atp$ for different
  ranges of $p_{T,\ttbar}$. The red and green curves show these ratios
  for events with $p_{T,\ttbar}<20\gev$ and $p_{T,\ttbar}>20\gev$,
  respectively. The blue curve is the ratio across all $p_{T,\ttbar}$
  values. Most of the cross section resides in the lower
  $p_{T,\ttbar}$ region, so the errors on the high-$p_{T,\ttbar}$
  region are larger because of limited statistics. All calculations
  were performed in the idealized SM scenario, at NLO in top quark
  pair production.}
\end{figure}

\section{Realistic case: SM}\label{sec:realistic}

We have seen that, in an ideal detector, the lepton and $\ttbar$
asymmetries are correlated and follow a robust, predictable curve as
a function of the lepton $p_T$. We must now show to what extent this
correlation remains intact in a true hadron collider environment. We
proceed in two steps. First, we continue with a parton-level analysis,
but impose a set of realistic cuts employed but the collaborations in
the actual analysis; for concreteness we are going to apply the cuts
used by the CDF collaboration, the ones used by the D\O\ collaboration
are in practice very similar and have negligible impact on our final
conclusions. Second, to further close the gap with the true
experimental conditions, we repeat our study including parton shower
effects and genuine top quark reconstruction.

\subsection{Parton-level analysis including cuts}
\label{sec:cuts}

Including possible real emission, the parton-level process for
$\ttbar$\/ production at NLO has 7 final-state particles: one lepton,
one neutrino, and up to 5 jets, two of which originate from $b$\/ or
$\bar b$\/ quarks. Inspired by CDF~\cite{Aaltonen:2011kc}, we impose
the following cuts on these objects. We require:
\begin{itemize}\setlength{\itemsep}{4pt}%\setlength{\parskip}{0pt}
\item exactly one lepton of $p^l_T>20\gev$ and $|\eta^l|<1.1$\,.
\item $\slashed E_T>20\gev$, which we take directly from the $p_T$ of
  the neutrino.
\item jets to have $p_{T,\,\mrm{jet}}>20\gev$ and $|\eta_\mrm{jet}|<2.0$,
  and to be formed with the $k_T$ algorithm using a jet size of
  $R=0.4$\,.\footnote{In the CDF analysis the CDFmidpoint cone
  algorithm is used, not $k_T$.}
\item in addition to the above requirements, that all $b$\/ and $\bar b$\/
 jets are restricted to $|\eta^b|<1.0$\,. This treatment of bottom
 quarks means we are effectively tagging both the $b$\/ and $\bar b$\/
 quarks, whereas CDF and D\O\ only require one tagged jet. This small
 difference in acceptance may lead to a difference in the absolute
 values of our calculated asymmetries compared to experiment, but we
 do not expect it to change our main result.
\item isolation criteria to be satisfied:
  $\upDelta R_{\mrm{jet},\,\mrm{jet}}>0.4$ and $\upDelta R_{l,\,\mrm{jet}}>0.4$\,.
\end{itemize}

Keeping the jet and missing energy cuts fixed to the above, we
calculate $\Atp$ and $\Alp$, just as we did in the ideal case. As a
first step to evaluate this asymmetry, and in particular to evaluate
$\Atp$, we take the top quark truth information directly from the
simulation (no reconstruction). The result is shown below in
Fig.~\ref{fig:cdfcuts}. While the absolute value of the asymmetries is
diminished once cuts are imposed (see Fig.~\ref{fig:ideal} for a
comparison), the trend in $\Al$, $\At$ versus $\ptl$ is preserved.

\begin{figure}[t!]
\centering\vskip4mm
\includegraphics[clip,width=0.44\textwidth]{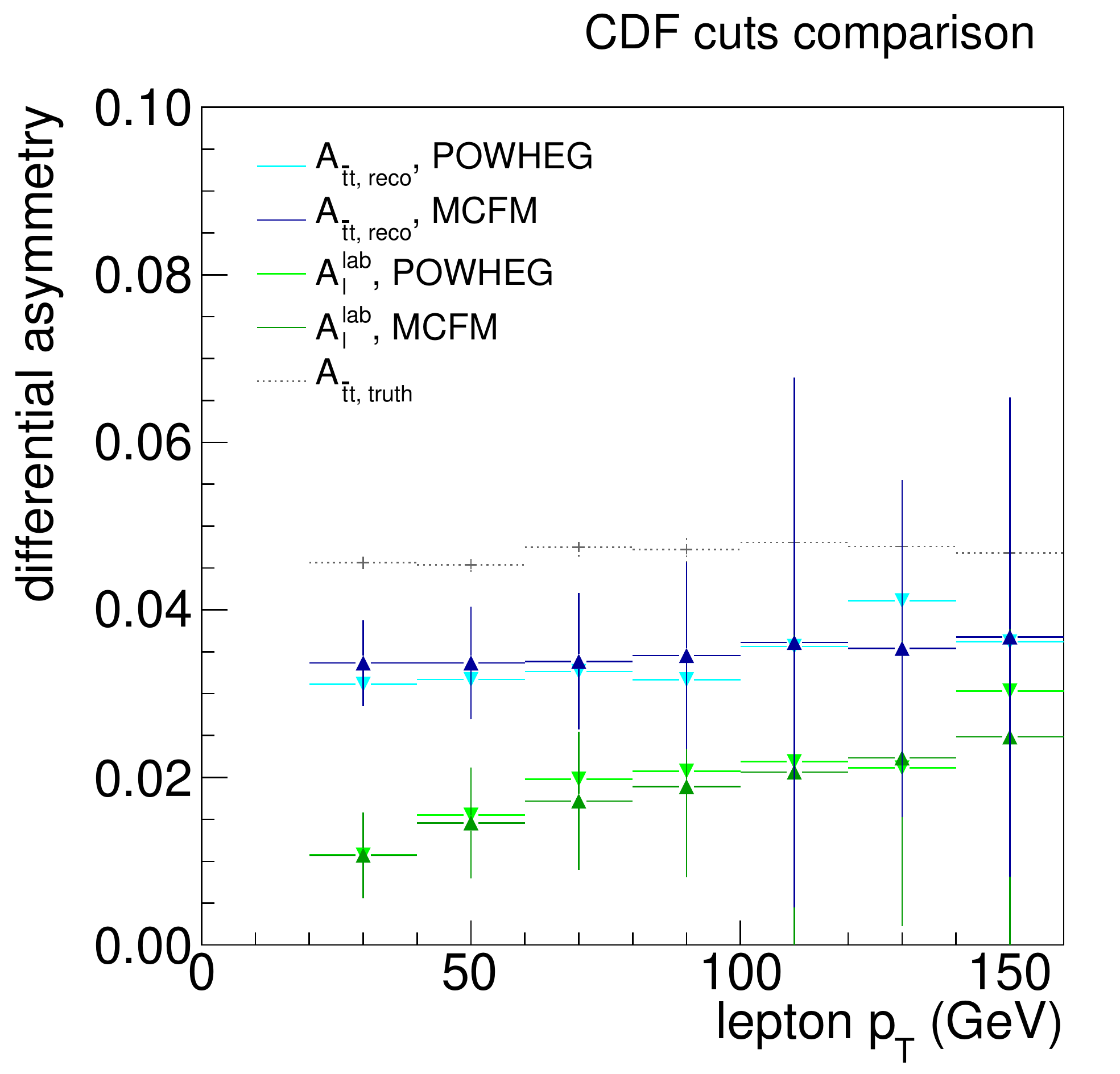}
\includegraphics[clip,width=0.44\textwidth]{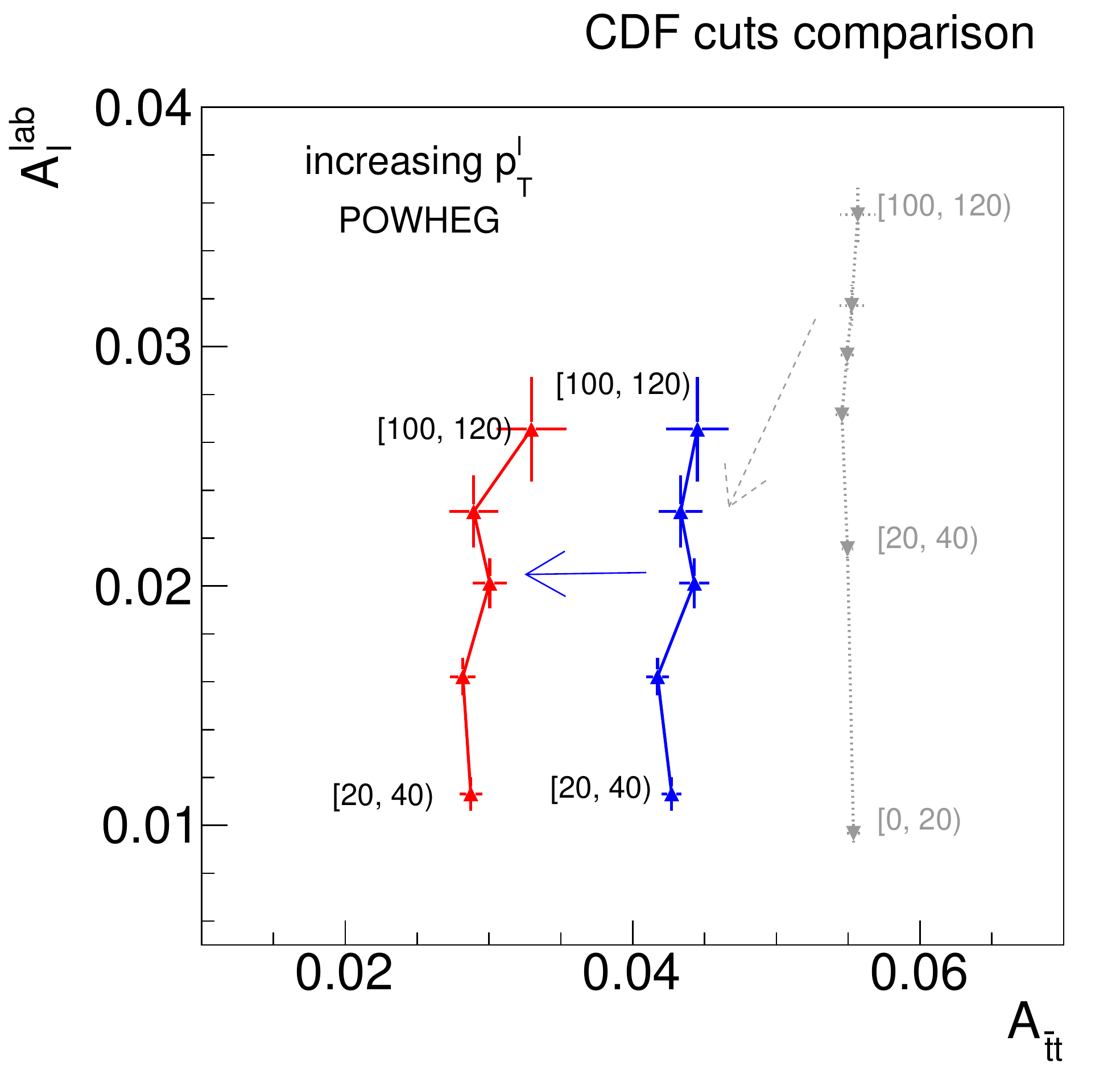} 
\caption{\label{fig:cdfcuts}
  Left: differential asymmetries $\Atp$ and $\Alp$, where all CDF
  analysis cuts have been imposed. The darker shaded lines indicate
  results obtained via \mcfm, while the lighter ones show the \powheg
  results. Both $\At$ and $\Al$ were calculated in the lab frame, with
  $\Al$ depicted in green. The blue lines show $\At$ calculated using
  reconstructed (anti)top quarks, following the scheme in
  Ref.~\cite{Campbell:2012uf}, rather than using the truth information.
  The grey line shows $\At$ including cuts, but calculated using the
  true top quark four-vectors rather than using reconstructed objects.
  Right: the $\Al\,(\ptl)[\At\,(\ptl)]$ post-cuts curve with (red
  line) and without (blue line) reconstruction compared to the
  $\Al[\At]$ curve obtained in the ideal case (grey). Because of the
  lepton cut imposed in the CDF analysis, the new
  $\Al\,(\ptl)[\At\,(\ptl)]$ curve starts at $20\gev$ rather than
  zero. The brackets indicate the $\ptl$ range corresponding to
  particular points along the $\Al\,(\ptl)[\At\,(\ptl)]$ curve. The
  grey arrow indicates how the $\Al,(\ptl)[\At\,(\ptl)]$ curve shifts
  from cut effects, while the blue arrow shows the effect of
  reconstruction. Note that only $\At$ is affected by the
  reconstruction, while both asymmetries are decreased by the analysis
  cuts.}
\end{figure}

Another step towards a more realistic analysis is to redo our top
quark analysis in terms of {\em reconstructed}\/ objects rather than
using the truth partonic information (before the decay). While a
completely realistic reconstruction will be presented in the next
section and will include parton showering effects, we can study some
reconstruction effects even at the parton level. Specifically, even at
the parton level, jets are sometimes lost due to acceptance or gained
from ISR (initial-state radiation), leading to an incorrect
reconstruction and a warped $\At$. To study this effect, we compare
the $\At$ calculated using perfect top quark reconstruction with the
$\At$ calculated using a reconstruction where all 5 jets, not just those
coming from the top quark decays, are considered.\footnote{Specifically,
  we use the ``improved'' reconstruction introduced in
  Ref.~\cite{Campbell:2012uf}: all non-$b$\/ jets are considered in
  the hadronic $W$\/ and $t$\/ reconstruction, and the combination (2
  {\em or}\/ 3 partons) that has invariant mass closest to $m_W$ or
  $m_t$ is selected.}
This comparison is shown in Fig.~\ref{fig:cdfcuts} as well. The impact
it has on $\At$ is sizable, amounting to a ${\cal O}(\text{30\%})$
reduction in $\At$. We find that both effects, namely including the
acceptance cuts as well as the top quark reconstruction reduce the
resulting asymmetries. This is expected since they force the events to
be more central, and wrong partonic assignment in realistic
reconstruction dilutes the asymmetry.

Finally, we have repeated the scale variation check with the post-cut
parton-level events and find the same trend as in the ideal cuts case
(cf.\ Fig.~\ref{fig:scalevar}): while the individual asymmetries
$\Atp$ and $\Alp$ shift by ${\cal O}(\text{25--30\%})$ as the
factorization/renormalization scale is varied by a factor of two, the
ratio $\Alp/\Atp$ remains stable.

\subsection{Showering and reconstruction effects:
  \powheg + \pythia and \sherpa's\\\cssr}
\label{sec:reco}

Having included cuts in our analysis and observed that the correlation
between $\At$ and $\Al$ is maintained, the next step towards reality
is to include a parton shower. The radiation, which quarks and gluons
emit as they lose energy in their evolution from the scale of the hard
process to the hadronization scale can show up as additional jets in
the detector. This spray of energy, and the combinatorial problem it
creates in any analysis relying on reconstruction, tends to dilute
parton-level effects. It is therefore important for us to show that
our correlation remains visible in the actual environment where the
experiments reside.

\begin{figure}[t!]
\centering\vskip0mm
\includegraphics[clip, width=0.44\textwidth]{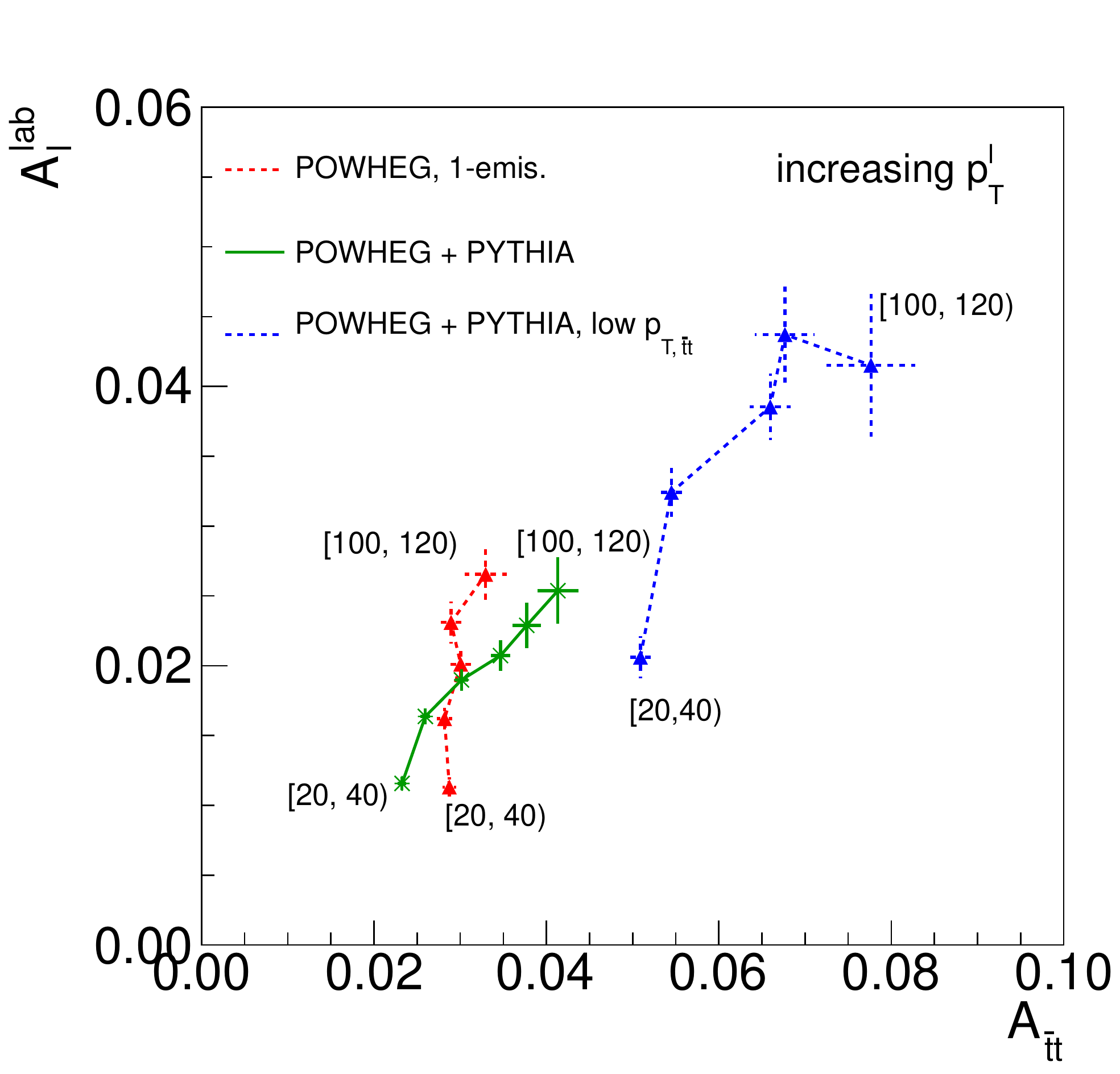}
\includegraphics[clip, width=0.44\textwidth]{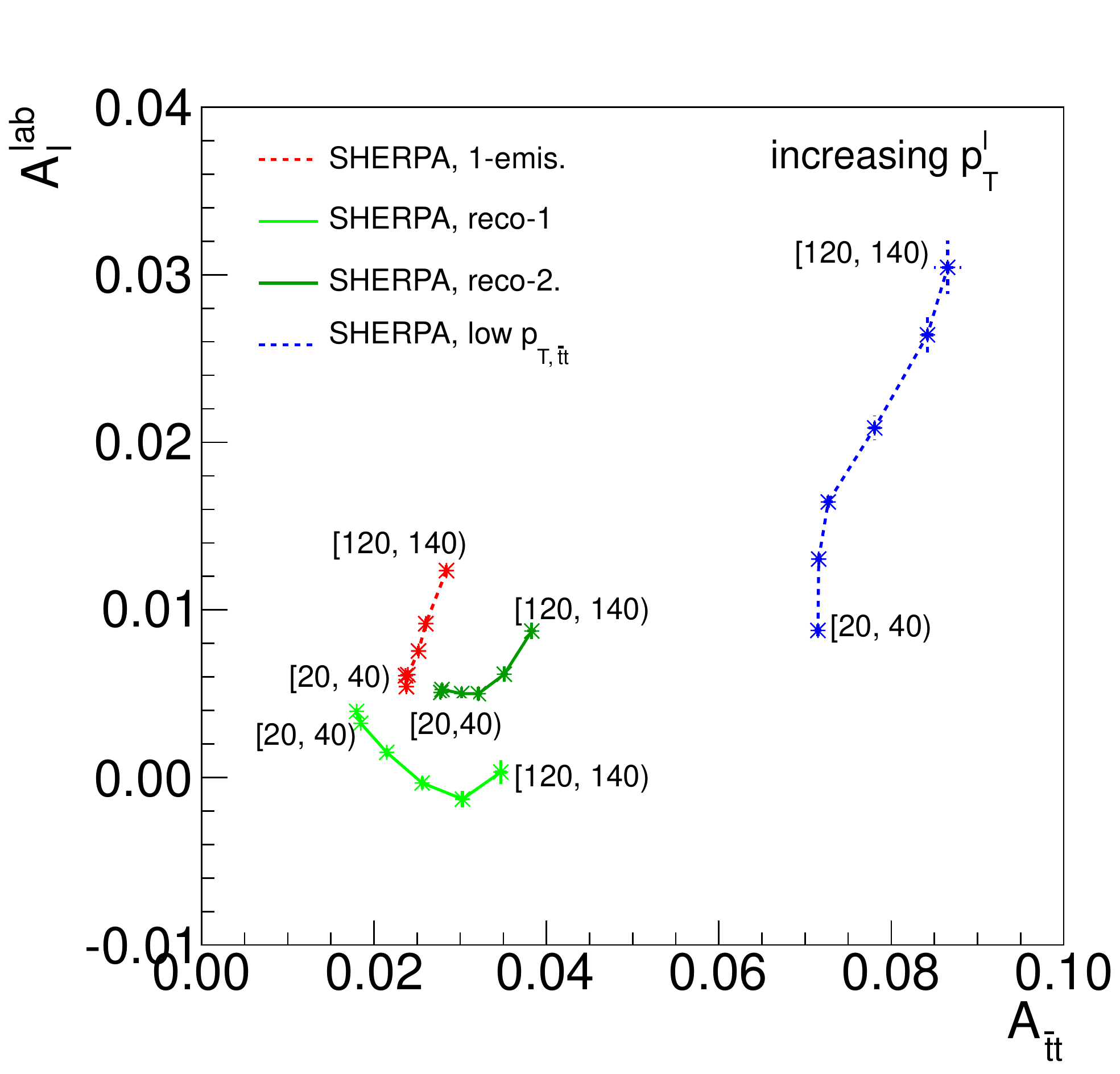}
\caption{\label{fig:powhegpythia}
  Left: the $\ptl$ dependence of the leptonic and $\ttbar$\/
  asymmetries for \powheg events that have been showered and
  hadronized with \pythia. The post-shower $\Al\,(\ptl)[\At\,(\ptl)]$
  curve is shown in dark green and is overlayed on top of the result
  with cuts and reconstruction done at parton level (red). The blue
  line shows the post-shower result in events where the total
  $p_{T,\ttbar}$ is restricted to $p_{T,\ttbar}<10\gev$. Right: the
  analogous curves for events generated with \sherpa's \cssr. The red
  curve shows $\Al\,(\ptl)[\At\,(\ptl)]$ with the \cssr truncated at
  the first emission, the blue curve shows the fully showered result
  for events restricted to $p_{T,\ttbar}<10\gev$, and the green curves
  depict the inclusive, fully showered \sherpa results. The difference
  between the green curves is the reconstruction algorithm. In the
  dark green curve, as described in the text, the number of jets
  included in the reconstruction algorithm is allowed to float, while
  in the light green curve the number is restricted to four, giving a
  straightforward, unambiguous reconstruction out of the hardest $b$\/
  and $\bar b$\/ jet and the two leading light-flavour jets. The
  low-$p_{T,\ttbar}$ result uses the latter reconstruction algorithm.}
\end{figure}

One way to study the effects of the parton shower on our observable is
to pass NLO \powheg events through \pythia~\cite{Sjostrand:2007gs}.
Manipulating the particle-level \pythia output into jets via
\fjet~\cite{Cacciari:2005hq,Cacciari:2011ma},\footnote{We use the
  $k_T$ algorithm with jet size $R=0.4$. Jets are identified as $b$\/
  jets if there is a $b$ ($\bar b$) parton within the jet area.}
we can then apply the same analysis cuts as in the previous section.
The only subtlety is how we handle $b$-tagging. In the simple
reconstruction used in the last section, we assumed the identity of
both $b$\/ quarks was known, leaving no ambiguity about whether/how to
combine the tagged jets with the lepton. In this section, we drop this
assumption, choosing to select and reconstruct events exactly as in
the CDF analysis~\cite{AFBCDF1}. Specifically, provided an event
contains a lepton and missing energy passing criteria shown in
Sec.~\ref{sec:cuts}, the event is kept if it contains four or more
jets and {\em at least one}\/ $b$-tag. The leading four jets, one of
which {\em must}\/ be a $b$\/ jet, are subsequently divided into a
hadronic top quark system and one jet to be paired with the lepton +
$\slashed E_T$ system. Each combinatoric possibility in the leptonic
$W$\/ reconstruction and the division of jets is tested, and the
combination that best reconstructs the top quark masses (the hadronic
and leptonic) is retained. The differential asymmetries for the
reconstructed top quarks (and leptons) are shown in
Fig.~\ref{fig:difflevels} below, and in the $\Al\,(\ptl)[\At\,(\ptl)]$
plane on the left panel of Fig.~\ref{fig:powhegpythia}. For
comparison, we also show $\Atp$ and $\Alp$ at parton level.

\begin{figure}[t!]
\centering\vskip0mm
\includegraphics[clip,width=0.44\textwidth]{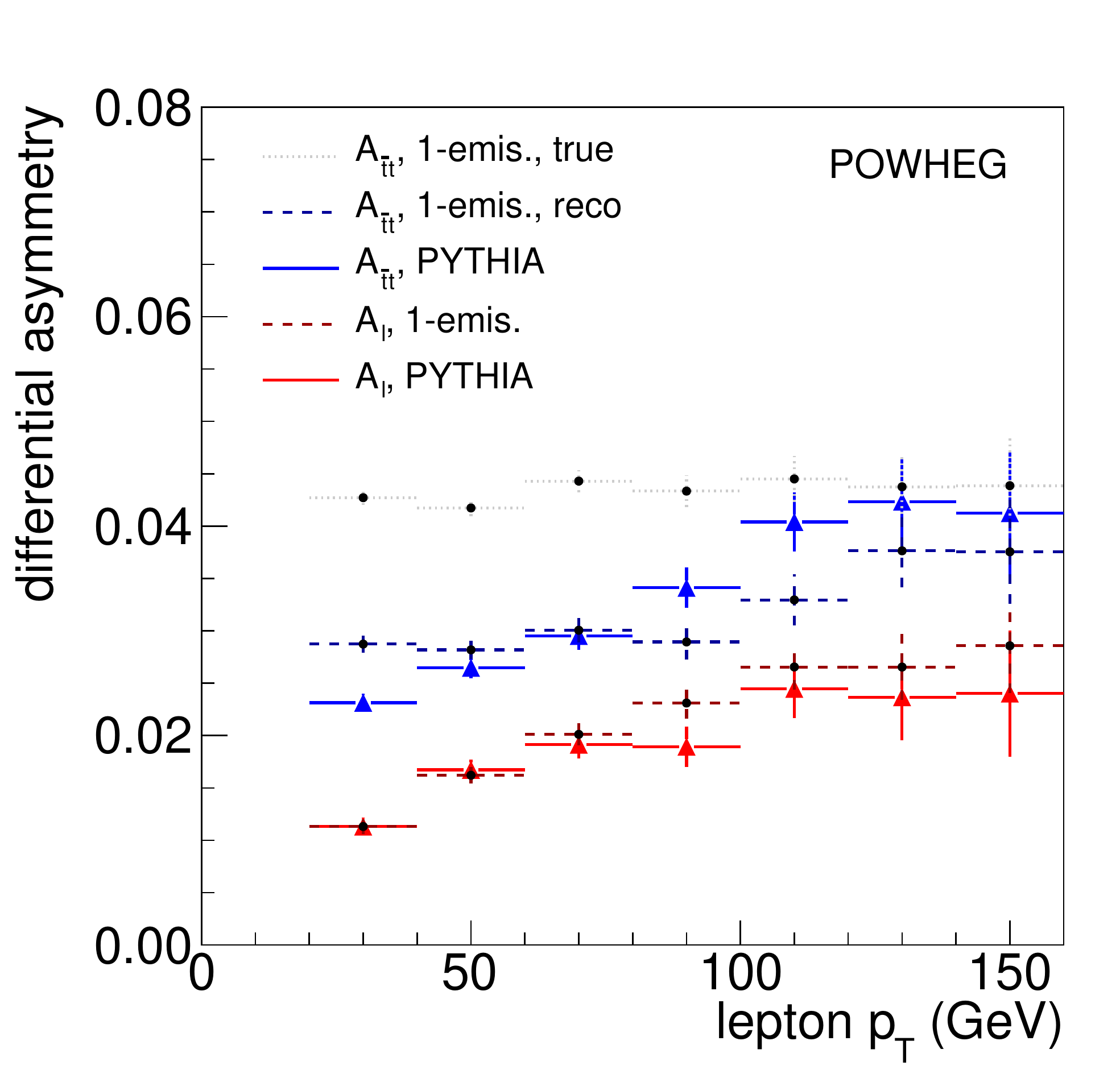}
\caption{\label{fig:difflevels}
  Dependence of the differential top quark and lepton asymmetries,
  versus $\ptl$, on parton shower and reconstruction effects. The truth
  values of $\Atp$ -- after cuts, but without any reconstruction or
  showering effects -- are shown in the uppermost line (grey, dotted).
  Including reconstruction but staying at parton level, $\Atp$ shifts
  to the dashed blue line. Finally, the post-shower, post-reconstruction
  values are given in solid blue. Similarly, the post-\pythia, lab
  frame $\Alp$ is shown by the red solid line, with the red dashed
  line indicating the lab frame $\Alp$ at parton level.}
\end{figure}

The experiments typically compare asymmetries at the
background-subtracted and unfolded levels, the analog of comparing
our fully showered and ``ideal scenario'' results. However, as we are
dealing with Monte Carlo data rather than experimental data, we are
able to break down these shifts in multiple stages. Being more
quantitative (cf.\ Fig.~\ref{fig:cdfcuts}), we have seen that imposing
cuts reduces the leptonic asymmetry by roughly 50\% and $\At$ by 20\%
when compared to the ``ideal case'' values. Reconstruction does not
affect $\Al$, but it further reduces $\At$ by an additional 30\%.
Working at the parton level, these reductions are almost independent
of the lepton $p_T$, our proxy for the energy of the $\ttbar$ system.
Showering and more realistic reconstruction have a $\ptl$ dependent
effect; $\Al$ is shifted by ${\cal O}(\text{1--10\%})$, and $\At$ is
shifted by ${\cal O}(\text{20--30\%})$, with the shifts increasing
with $\ptl$.

As a second study of the effects of a parton shower on the $\Al-\At$
relationship, we analyze the asymmetry correlation using
\sherpa (v1.4.0)~\cite{Gleisberg:2008ta}. Despite the fact that
\sherpa is a LO matrix element generator, the addition of a parton
shower technique that appropriately includes color-coherence effects
will generate a forward--backward top quark asymmetry (see
Ref.~\cite{Skands:2012mm}), so it is interesting to see what degree it
retains the $\Al-\At$ correlation. To generate events, the LO matrix
elements, including the decays of the top quarks, are showered
according to \sherpa's color-coherent showering (\cssr)
description~\cite{Schumann:2007mg,Hoeche:2009xc}. In the infrared
limit of QCD, this description correctly accounts for ISR effects,
intermediate top quark radiation and multiple emissions from the
final-state $b$\/ quarks and decay partons.\footnote{For the hard
  process generation, we use CTEQ6L1 PDFs \cite{Pumplin:2002vw} and an
  $m_{\perp,t}$-like scale choice ($m^2_{\perp,t}=m^2_t+(p^t_T)^2$)
  resulting from utilizing the default scale setting prescription
  applied in \sherpa. The top quark decays are incorporated at full
  matrix element level, \ie preserving spin correlations and full
  width effects beyond the narrow-width approximation, with the only
  requirement of producing intermediate top quark states
  ($m_t=173.2\gev$).}
While qualitative agreement in $\At$ between the LO matrix element
plus parton shower (LO + PS) results and the complete NLO results has
been demonstrated \cite{Skands:2012mm}, exploring $\Al-\At$ tests the
adequacy of the LO + PS calculation on a detailed and differential
level.

The \sherpa events are reconstructed in similar, but slightly
different manner than we used in the \powheg + \pythia case. In
particular, the charge of the $b$\/ jets is assumed to be known,
leading to perfect assignment of the $b$\/ ($\bar b$) quarks to the
top (antitop) quark.\footnote{We obtain this information from a $k_T$
  jet algorithm where we systematically keep track of the $b$-parton
  flavours.}
After combining the lepton and missing energy with the correct $b$\/
jet, the top quark objects are reconstructed by testing {\em all}\/
possible partitions of jets -- even beyond the 4 jets required for
event selection -- and keeping the combination which yields the
smallest summed-up mass deviation, which we take as
$|m^\mrm{lep}_{\trm{pseudo-}t}-m_t|+|m^\mrm{had}_{\trm{pseudo-}t}-m_t|+
\sqrt{2}\,|m^\mrm{had}_{\trm{pseudo-}t,\mrm{jets}}-m_W|$. By including
more jets, the combinatorial issues are bigger for this reconstruction
method, however it will capture situations the CDF method cannot, \ie
where the hadronic top quark radiates and manifests itself actually a
4-jet system, or the jet system of the leptonic top quark possesses an
extra jet beyond the $b$\/ jet. The leptonic and $\ttbar$\/
asymmetries derived from the reconstructed \sherpa events are shown in
the right panel of Fig.~\ref{fig:powhegpythia}. We also show the
asymmetries obtained from running \sherpa's \cssr in ``one-emission''
mode, where showering is terminated after no more than one emission
has occurred. The one-emission mode allows us to see the impact of the
multi-parton emissions and the (enhanced) combinatorial headache
coming with them.
 
Inspecting Fig.~\ref{fig:powhegpythia}, while the shapes of the
differential asymmetry curves at parton (or ``one-emission'') level
agree reasonably, the results after the full shower are quite
different. The slope of the \powheg + \pythia curve acquires a slight
tilt, as the later/secondary shower emissions drive the slope of
$\Atp$ and $\Alp$ slightly in opposite directions (see
Fig.~\ref{fig:difflevels}). For the \sherpa result, higher/secondary
emissions drive $\At$ and $\Al$ in the same directions as \powheg +
\pythia (the $\Al$ is slightly drecreased, while the $\At$ is
enhanced,\footnote{A larger enhancement from \sherpa's \cssr is
  expected~\cite{Skands:2012mm} because subsequent emissions in the
  shower are still carried out in a color-coherent manner.}
especially at high $\ptl$), however the change in slope is much more
dramatic: $\Al$ is decreased nearly to zero, or even negative, until
$\ptl\gtrsim!00\gev$. Massaging the \sherpa top reconstruction
algorithm does not fix the discrepancy -- the simplified
reconstruction used in the lower curve of the \sherpa panel in
Fig.~\ref{fig:powhegpythia} is closer to the method used for \powheg,
yet the $\Al\,(\ptl)[\At\,(\ptl)]$ curve is just as disparate.

The low-$p_{T,\ttbar}$ curves in Fig.~\ref{fig:powhegpythia} offer
some insight into why the post-shower results disagree. The transverse
momentum of the $\ttbar$ system separates the so-called ``Sudakov
region'' -- where $\At$ is positive --  from the ``hard-$p_T$ region''
where the entire $\ttbar$ system recoils against additional radiation
and the asymmetry is negative. By imposing a low cut on $p_{T,\ttbar}$
we can focus on the effects that drive the positive asymmetry and are
generated by the virtual corrections and multiple soft gluon emission.
The majority of the cross section resides in this low-$p_T$ region, so
it is important to understand and study the asymmetry here. As we can
see, the low $p_{T,\ttbar}$ curves closely follow the ideal NLO case
(cf.\ the discussion in Sec.~\ref{sec:ideal}), despite the fact that
all virtual pieces beyond the soft and collinear approximation, \ie
beyond the lowest order Sudakov description are absent in the \sherpa
\cssr description. In other words, for regions of phase space close to
Born kinematics, the LO + PS and NLO calculations are in relative
agreement. Note that the denominator in the \sherpa asymmetry
calculation is the lowest order cross section and is smaller than the
NLO cross section used for \powheg.

However, turning to the inclusive case, the calculation becomes
sensitive to the whole $p_{T,\ttbar}$ region and how it is modeled. As
discussed in~\cite{Skands:2012mm}, in color-coherent parton showers
like the \cssr, the real-emission effect is overestimated because
leading-$N_C$ color factors, rather than the smaller, NLO-correct
factors, are utilized and the use of the eikonal limit/dipole
radiation functions is extended beyond its/their reliable range. The
enhanced emission leads to a stronger reduction of the already smaller
virtual effects as soon as we go away from the low $p_{T,\ttbar}$
limit. The shower Sudakov effect (the enhanced asymmetry in the
low-$p_{T,\ttbar}$ region) is weaker than the enhancement in the full
NLO calculation, so the negative contributions from emission win out
and drag the asymmetries down. So, although the \sherpa \cssr
qualitatively generates an $A_{\ttbar}\,(p_{T, \ttbar})$ similarly to
\powheg + \pythia, it cannot reproduce it in the details, and the
details are important to get the $\Al\,(\ptl)[\At\,(\ptl)]$
correlation right. One reason the lepton asymmetry may be more
sensitive to the shower details may be the tight rapidity cut,
$|\eta^l|<1.1$, whereas $|\eta_\mrm{jet}|<2.0$\,. Being so central,
the leptons can easily be nudged from forwards-moving to
backwards-moving (or vice versa) or driven out of (into) the
acceptance range, making $\Al$ more susceptible to later shower
emissions. We emphasize that the instability shown in
Fig.~\ref{fig:powhegpythia} is not \sherpa-specific, any LO + PS
calculator will struggle to capture the details of the
$\Al\,(\ptl)[\At\,(\ptl)]$ correlation for the reasons mentioned
above.

To summarize, we find that the qualitative features of the SM
$\Atp-\Alp$ correlation are maintained throughout all levels of event
and analysis complexity. The absolute values of the asymmetries and
their relation to each other do shift depending on the stage of the
analysis, but the slope of the $\Al\,(\ptl)[\At\,(\ptl)]$ curve is
maintained. Importantly, to see the robustness of this result, one
{\em must}\/ use calculational tools that are exact at NLO (such as
\powheg + \pythia). Lowest order matrix element calculators
supplemented with a parton shower respecting color coherence may be
sufficient to describe gross features of $\At$ and do offer valuable
insight to the physics behind the asymmetry, but these tools lack some
of the physics necessary to capture detailed effects like the
correlation between differential asymmetries. While the sensitivity of
differential correlations shown here should serve as a warning label
on calculations done with LO + PS accuracy, as we will show in detail
in the next section, the $\Al\,(\ptl)[\At\,(\ptl)]$ curves in various
benchmark BSM scenarios are dramatically different from the SM curve.
Therefore, small shifts (in absolute value) in the asymmetries
originating from cut, reconstruction or shower effects will not
seriously hinder the discriminating power of the
$\Al\,(\ptl)[\At\,(\ptl)]$ correlation.

%%%%%%%%%%%%%%%%%%%%%%%%%%%%%%%%%%%%%%%%%%%%%
\section{Asymmetries beyond the standard model}\label{sec:BSM}

As we have discussed, in the SM the shape of $\Alp[\Atp]$ can be
simply understood as a consequence of the fact that $t\bar t$\/
production in QCD is {\em unpolarized}.
In particular, the same number of left- and right-handed top quarks is
produced, and that equal amounts of $t\bar t$\/ pairs are produced in
collisions of left- and right-handed quarks.
The latter fact ensures that the lepton asymmetry vanishes at the
$t\bar t$\/ production threshold \cite{Falkowski:2011zr}.
On the other hand, the fact that the top quarks have no overall
polarization ensures that, for top quarks produced with a significant
momentum, the lepton direction is determined by the kinematics of the
boosted top quark decay.
These simple arguments explain the behavior of $\Alp/\Atp$ in the SM.  
At small $\ptl$ the lepton asymmetry is dominated by the
near-threshold sample, and one expects $\Alp\ll\Atp$.
As $\ptl$ grows the lepton direction becomes more and more correlated
with that of the parent top quark, therefore one expects $\Alp$ to
approach the $\Atp$ curve from below. This is indeed what comes out of
the MC simulations we presented in Secs.~\ref{sec:ideal}~and~\ref{sec:realistic}.
Moreover, we found that the top quark asymmetry is approximately
independent of  $\ptl$ due to an interplay between two effects: the
increase of $\At$ as a function of $m_{t\bar t}$ (correlated with
$\ptl$), and the increase of $\At$ as a function of $\upDelta Y_{t\bar t}$
(anti-correlated with $\ptl$).

These expectations can be grossly violated in models beyond the SM.
New physics may introduce two effects that could potentially distort
the shape of $\Alp[\Atp]$ away from the SM prediction.
One is that BSM models addressing the anomalous $t\bar t$\/ asymmetry
at the Tevatron may lead to a very different $m_{t\bar t}$ dependence
of the asymmetry.
Due to the correlation between $\ptl$ and  $m_{t\bar t}$, in the
presence of new physics there is no reason for $\Atp$ to remain
approximately constant as in the SM. In fact, as we will see below,
$\Atp$ can either increase or decrease, and the strength of the effect
is strongly model dependent.
The other important effect is polarization. Models addressing the
anomalous top quark asymmetry measured at the Tevatron introduce new
particles with different couplings to left- and right-handed quarks.
This typically leads to polarization both in production and in the
final state. These polarization effects may be observable, as pointed
out many times in the literature~\cite{Godbole:2010kr,Cao:2010nw,Choudhury:2010cd,Barger:2011ya,Berger:2011hn,Berger:2011pu,Berger:2012tj}.
In particular, the inclusive~\cite{Krohn:2011tw} and threshold
\cite{Falkowski:2011zr} lepton asymmetries are important observables
to test the SM and discriminate between different models of new
physics predicting the same top quark asymmetry. Here we point out
that studying the $\ptl$ dependence of $\Atp$ and $\Alp$ provides
further discriminating power.

To illustrate it, we study $\Atp$ and $\Alp$ in BSM models that lead
to an enhanced top quark asymmetry in agreement with the Tevatron
observations. We focus on the so-called {\em s-channel axigluon}.
In this model one introduces a color-octet field $G_\mu^a$ with mass
$m_G$ and the couplings to quarks that are assumed to be flavor
diagonal but otherwise arbitrary:
\begin{equation}
\label{eq:axigluon}
{\cal L}\;\subset\;g_{L,i}\,\bar q_i\,\bar\sigma^\mu\,G_\mu^a\,T^a\,q_i\;+\;
g_{R,i}\,q_i^c\,\sigma^\mu\,G_\mu^a\,T^a\,\bar q_i^c\,,
\end{equation}
where $q_i$ and $q^c_i$ denote left-handed doublet and right-handed
singlet quarks, respectively. If the axigluon couplings to the top and
light quarks are chiral, $g_L\neq g_R$, then interference of the
$s$-channel axigluon exchange with the tree-level QCD gluon exchange
results in a non-zero forward--backward asymmetry of the $q\bar q\to
t\bar t$\/ production process.
There exist regions in the parameter space of the axigluon mass and
couplings leading to an ${\cal O}(\text{10\%})$ additional contribution
to the inclusive top quark asymmetry, bringing the theoretical
prediction to a good agreement with the central value observed by CDF
and D\O, while keeping the total $t\bar t$\/ cross section within the
experimental bounds.
To avoid the bounds from dijet production and $t\bar t$\/ resonance
searches, the axigluon should have a significant width,
$\mrm{\Gamma}_G\gtrsim0.2\,m_G$.

In the following we study in detail the predictions concerning $\Atp$
and $\Alp$ for a representative set of axigluon benchmark models.
One class has a relatively light and wide axigluon, $m_G=200\gev$,
$\mrm{\Gamma}_G=50\gev~{\rm GeV}$, with flavor universal couplings to
the SM quarks. For this mass, we pick three benchmarks with
left-handed, right-handed, and axial axigluon couplings to the quarks,
each predicting the same $\upDelta\At=0.12$ contribution to the
inclusive top quark asymmetry (on top of the SM contribution of
${\cal O}(\text{9\%})$), but each predicting a different lepton
asymmetry. The couplings, the axigluon width, and the computed lepton
asymmetries take the following values:
\begin{align}
\label{eq:axi200}
&{\rm(L)} \qquad g_{R,i} =  0,\, g_{L,i} = 0.8\, g_s\,:\quad  \, \upDelta \Al  = - 0.07, 
\nonumber \\
&{\rm(R)} \qquad g_{R,i} = 0.8\, g_s,\, g_{L,i} = 0\,:\quad  \, \upDelta \Al  = 0.18, 
\nonumber \\
&{\rm(A)} \qquad  g_{R,i} = 0.4\,g_s,\, g_{L,i} = -0.4 g_s\,: \quad \, \upDelta \Al  = 0.05, 
\end{align}
where $g_s$ is the QCD coupling. For those benchmarks the axigluon decay width into $q \bar q$ pairs is of order a GeV, thus the larger width we assumed must be explained by exotic decay channels (see \eg \cite{Tavares:2011zg}). 

The other class of benchmarks has a relatively heavy axigluon, $m_G = 1.5$~TeV, and flavor non-universal couplings 
(see~\cite{Frampton:1987dn, Djouadi:2009nb, Bauer:2010iq, Delaunay:2010dw, Bai:2011ed, Barcelo:2011fw, Barcelo:2011vk, Delaunay:2011vv, DaRold:2012sz } for some theoretical construction of heavy axigluon models).  
Again we pick 3 benchmarks with left-handed, right-handed, and axial axigluon couplings:
\begin{align}
\label{eq:axi1500}
&{\rm(L)} \quad g_{L,q}=-1.3\,g_s,\ g_{R,q}=0,\ g_{L,t}=6\,g_s,\ g_{R,t}=0\ :\
\upDelta A_{l}=-0.01,\ \mrm{\Gamma}_G=970\gev\,,\nn\\
&{\rm(R)} \quad g_{L,q} = 0,\, g_{R,q} = -1.1 g_s,\, g_{L,t} = 0,\, g_{R,t} = 6 g_s: \, \upDelta A_{l}  =  0.14, \,\mrm{\Gamma}_G = 460~{\rm GeV}\\
&{\rm(A)} \quad  g_{L,q} = 0.6\,g_s, \ g_{R,q} = -0.6\, g_s, \ g_{L,t} = -3\,g_s, \ g_{R,t} = 3\,g_s:~ \ \upDelta A_{l}  =  0.06, \ \mrm{\Gamma}_G = 350~{\rm GeV}\nn
\end{align}
where above $q$ stands for doublet and up-type singlet quarks of the first and the second generation (the couplings to down-type singlet quarks are assumed to vanish). 
In this case the axigluon width is assumed to be dominated by 2-body decays into the SM quarks, as follows from the coupling values listed above 
Much as for the light axigluon, each of the  benchmarks in Eq.~(\ref{eq:axi1500}) predicts $\upDelta A_{t \bar t}  = 0.12$.  

\begin{figure}[t!]
\centering\vskip0mm
\includegraphics[clip,width=0.48\textwidth,angle=0]{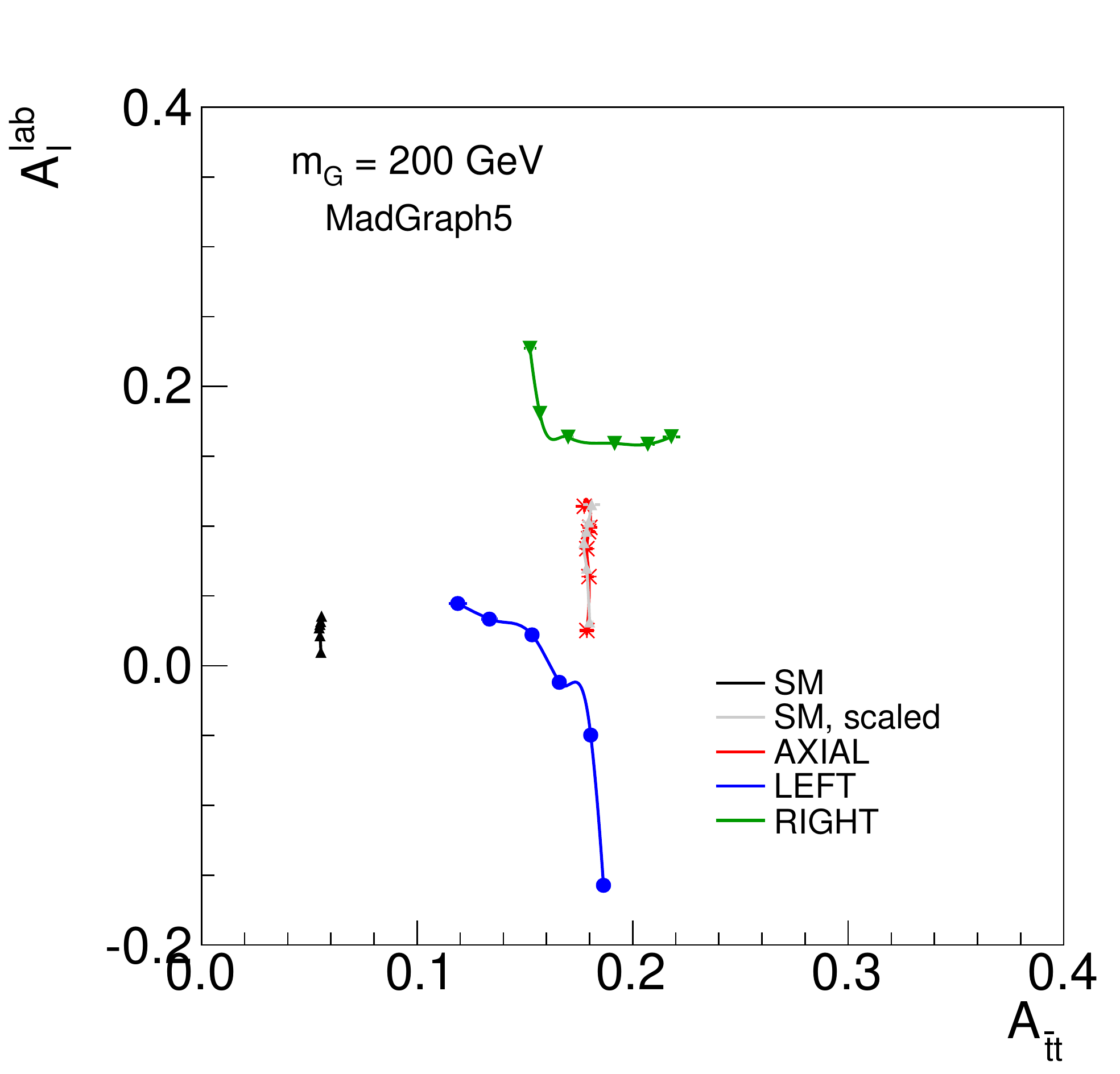}
\includegraphics[clip,width=0.48\textwidth,angle=0]{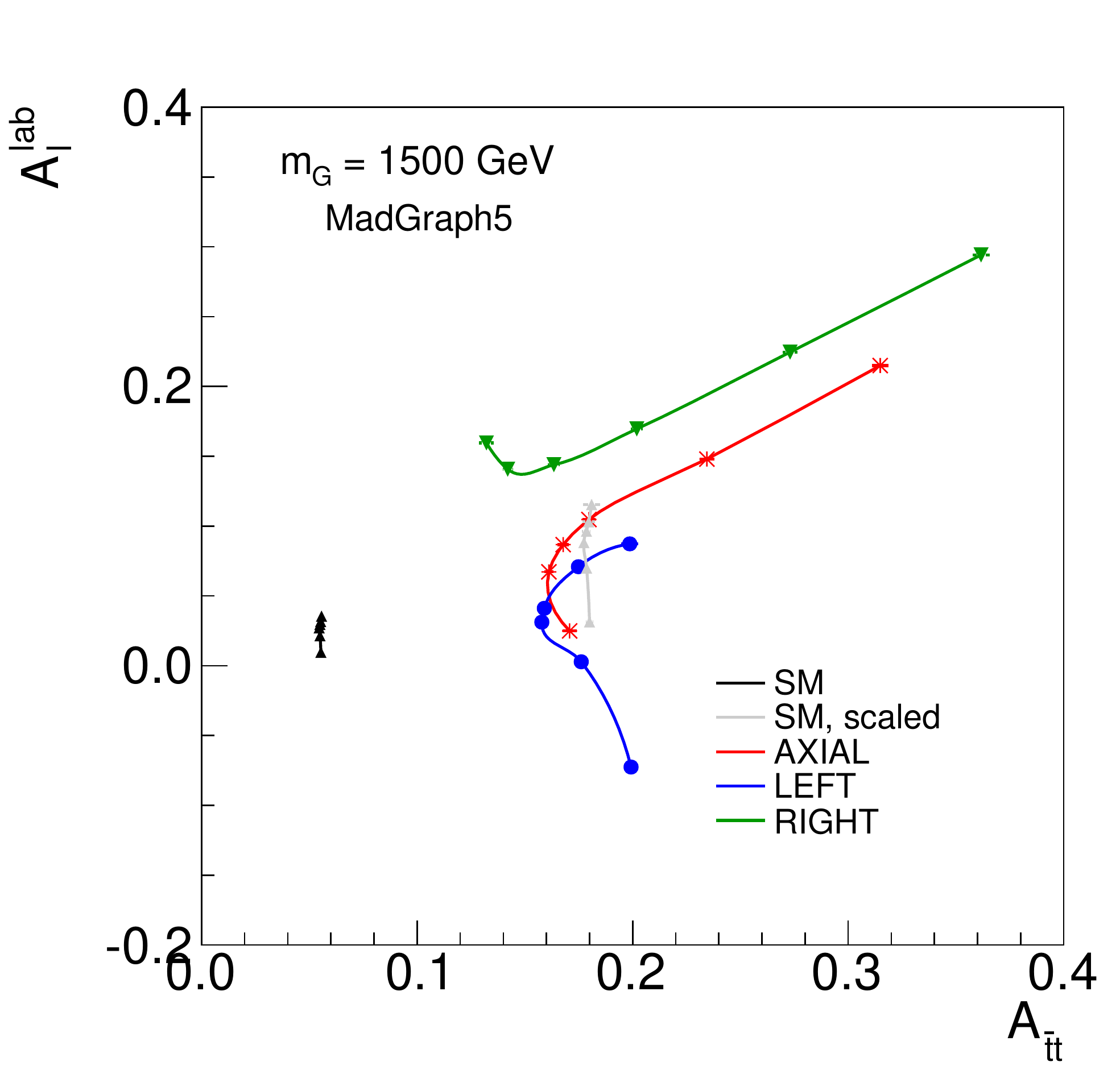}
\caption{\label{fig:bsm}
  Left: the $\Alp[\Atp]$ curves for increasing $\ptl$ bins in the
  axigluon benchmarks with $m_G=200\gev$ with left-handed (blue),
  right-handed (green), and axial (red) couplings, as specified in
  Eq.~(\ref{eq:axi200}). Also shown the analogous SM curve (black) for
  comparison, and a ``rescaled SM'' curve (grey), where $\At$ has been
  artificially inflated to $0.18$ -- the current experimental
  value. This rescaled curve emphasizes the difference in slope
  between the SM and our benchmark BSM scenarios. Right: the same
  curves presented for the axigluon benchmarks with $m_G=1.5\tev$ as
  specified in Eq.~(\ref{eq:axi1500}).}
\end{figure}

In Fig.~\ref{fig:bsm}, we plot the $\ptl$ dependence of the top and lepton asymmetries. 
To compute the asymmetries, we  simulated semileptonically decaying $t \bar t$ events using Madgraph 5~\cite{Alwall:2011uj} with a custom user-defined model describing the extension of the SM with the axigluon coupled to quarks as in  Eq.~(\ref{eq:axigluon}).
The asymmetries were computed using the parton level input without taking into account showering, hadronization, detector, or reconstruction effects. 
The samples were divided according to $\ptl$ into 6 bins with the lower bin limits at $0, 20, 40, 60, 100, 150$ GeV, and the points in Fig.~\ref{fig:bsm} refer to the asymmetries in these bins.   
As our simulations are tree-level and do not include the 1-loop SM contribution, we simply add to the results in each bin the SM asymmetries in that bin estimated by \powheg.\footnote{Both the SM and BSM asymmetries are dominated by interference of the tree-level QCD amplitude: with the 1-loop amplitude in the former case, and with the tree-level s-channel axigluon exchange in the latter, so it is reasonable to assume that the 2 effects add up.}   

As can be seen in Fig.~\ref{fig:bsm}, the differences  between the bechmarks become more pronounced when  $\ptl$ dependence  is explored.
The striking observation is that the slope of the curve traced in the  $A_{t \bar t}$-$A_{\ell}$ plane can be completely different than in the SM.  
For the light axigluon case,  the (R) benchmark predicts an approximately constant $\Alp$ and {\em increasing} $\Atp$, while for the (L) benchmark  $\Alp$ is increasing and $\Atp$ is {\em decreasing}. In both of these cases the shape of the curve can be qualitatively understood. At low $\ptl$ the lepton asymmetry is close to its thresholds value  and approaches $\Atp$ at large  $\ptl$. The shape of  $\Atp$ is influenced by top polarization: in the (L/R) case boosted top quarks have dominantly left/right helicity and tend to emit the lepton in the opposite/same direction as the top quark. For low and moderate $\ptl$ this leads to a certain degree of anti-correlation/correlation with $m_{t \bar t}$, and the derivative of $\Atp$ reflects the increasing $t \bar t$ asymmetry as function of $m_{t \bar t}$. 
On the other hand the (A) benchmark, where polarization effects are small and completely vanishing  at the $t \bar t$ threshold,  predicts the shape $\Atp$ and $\Alp$ similar to the SM, except that both lepton and top asymmetries in each bin are shifted to larger values. 

Very similar arguments hold for the heavy axigluon benchmarks. 
One important difference  between light and heavy axigluon benchmarks is that the latter predict a much steeper dependence of the top asymmetry on $m_{t \bar t}$. 
This is the reason why  $\ptl$ dependence is more pronounced for the heavy axigluon benchmarks. 
In  particular, $\Atp$ steeply grows at high $\ptl$ as a result of the correlation between  $\ptl$  and $m_{t \bar t}$. 
For the  (L) benchmark this leads to a  "turnaround"  of the curve  when anti-correlation between  $\ptl$  and $m_{t \bar t}$ at low $\ptl$ due to  top polarization turns into correlation at high $\ptl$. 

In summary, $\ptl$ dependence of the lepton and top asymmetries offers a handle to discriminate between the SM and new physics, and also between different models of new physics predicting the same top asymmetry. 
This test of the SM is to a large extent independent of the overall normalization of the asymmetries. 
If, hypothetically, some yet un-calculated higher-order QCD corrections to the asymmetries happen to be much larger than expected, they may shift the curve in the  $\Alp-\Atp$ to higher values without changing its shape.  
On the other hand, the shape of the curve in the  $\Al$ - $\At$  plane is very sensitive to polarization effects, which typically are present in BSM models addressing the anomalous top asymmetry at the Tevatron.

% ======= conclusions ==========================================================

\section{Discussion: additional reconstruction-independent quantities}
\label{sec:discussion}

Throughout this paper we have used $\ptl$ to study the relation
between $\Al$ and $\At$ in various kinematic regimes. The lepton $p_T$
is a useful variable to consider as a probe since it is simple, clean,
and -- just like $\Al$ -- it does not require reconstruction of any
complicated objects. However, it is not the only option. Along with
the properties of the lepton, there are several global observables,
such as the total invariant mass taken over all {\em visible}\/
final-state objects, $m_\mrm{vis}$, that are also reconstruction-independent.
A few other examples are $H_T$, $m_{T}$ and $m_{T,\mrm{vis}}$. Here,
$H_T$ is defined in the usual way as the scalar sum $H_T$ of {\em all}\/
identified-object $p_T$ including $\slashed E_T$; similarly, combining
all leptons and jets in the event (even beyond those required for
selection) into a single ``visible'' four-vector, $p_\mrm{vis}$, the
other transverse quantities are given as (cf.~Ref.~\cite{Lykken:2011uv}):
\begin{align}
m^2_{T}        &\;=\;
\left(\sum_{i=l,\,\mrm{jet}} |\vec{p}_{T,i}| + |\slashed E_T|\right)^2 -
\,\left(\vec{p}_{T, \mrm{vis}}+\vec{\slashed E}_{T}\right)^2\,,\nonumber\\[2mm]
m^2_{T, \mrm{vis}} &\;=\;\, m^2_\mrm{vis} + 2\,\big(
E_{T, \mrm{vis}}\,|\slashed E_T|\,-\,\vec{p}_{T, \mrm{vis}}\cdot\vec{\slashed E}_T
\big)\label{eq:tvars}
\end{align}
where
\begin{equation}
m^2_\mrm{vis}\;=\;p^2_\mrm{vis}\,,\quad
\vec{p}_{T, \mrm{vis}}\;=\;\sum_{i=l,\,\mrm{jet}} \vec{p}_{T,i}\,,\quad
E^2_{T, \mrm{vis}} = m^2_\mrm{vis} + p^2_{T, \mrm{vis}}
\end{equation}
denote the mass, transverse momentum, and transverse energy of the
visible system.

\begin{figure}[t!]
\centering\vskip-1mm
\includegraphics[clip, width=0.44\textwidth]{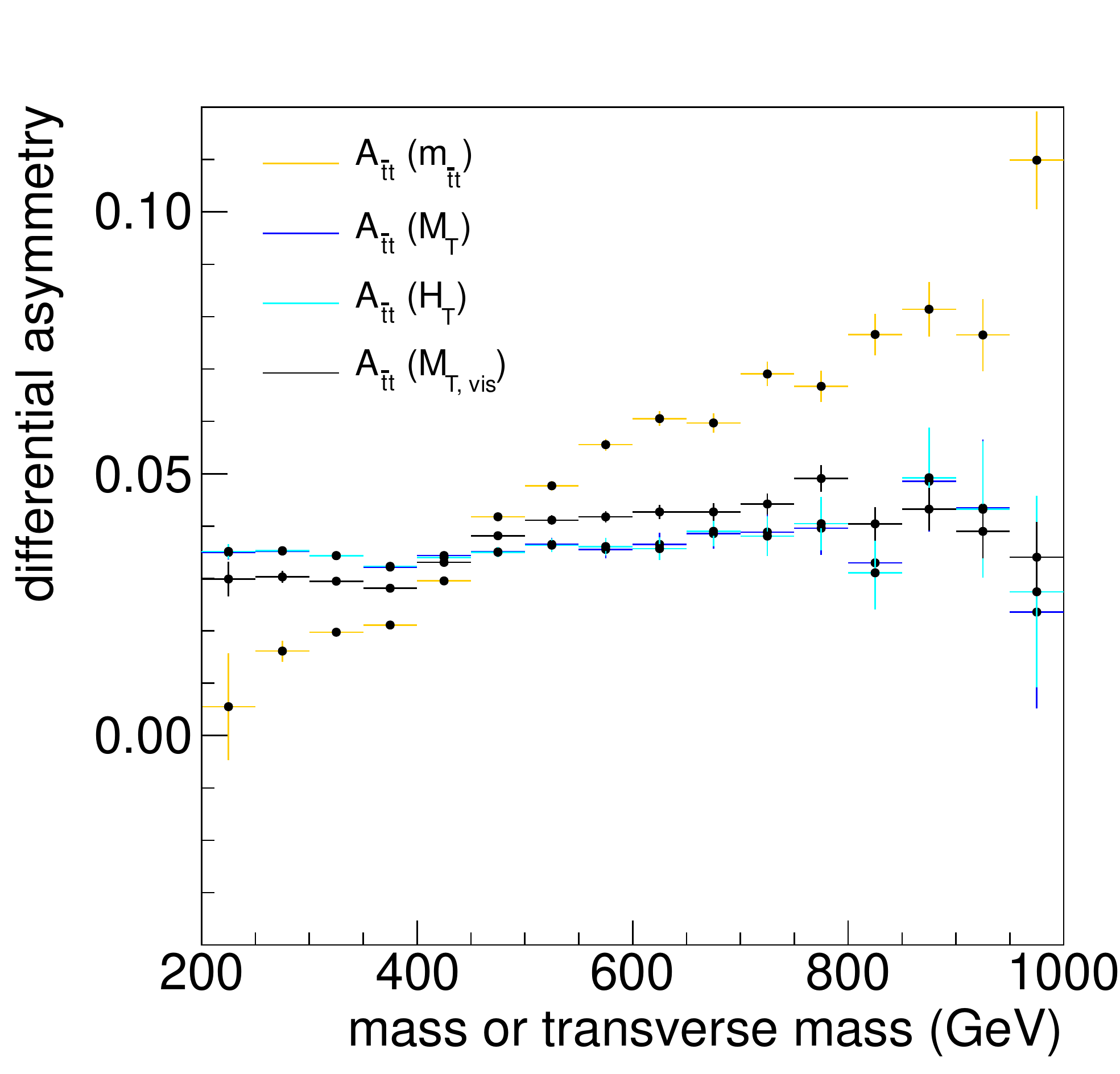}
\includegraphics[clip, width=0.44\textwidth]{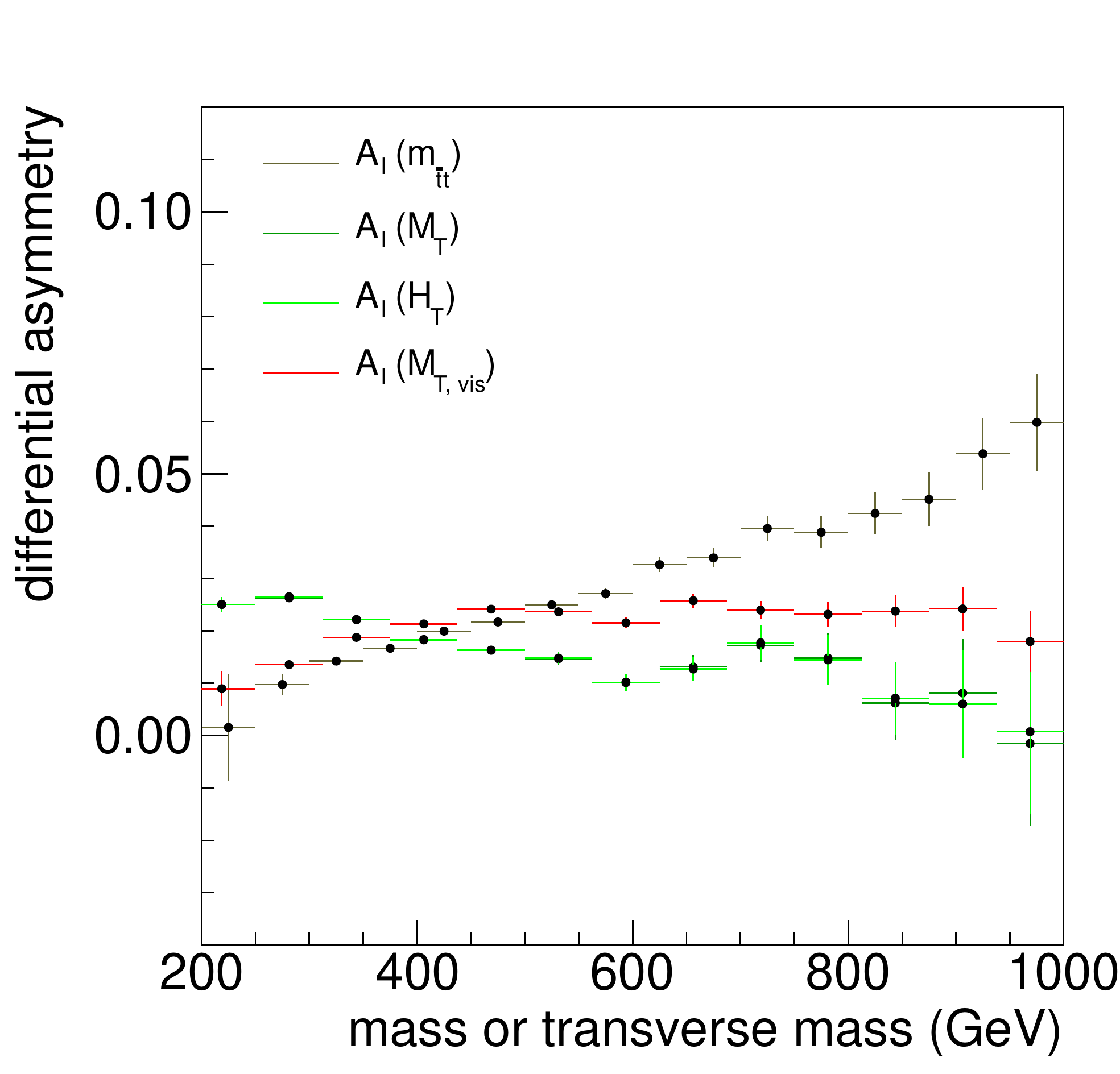}\\
\vskip5mm\hskip10mm
\includegraphics[clip,width=0.87\textwidth,angle=0, height=0.45\textwidth]{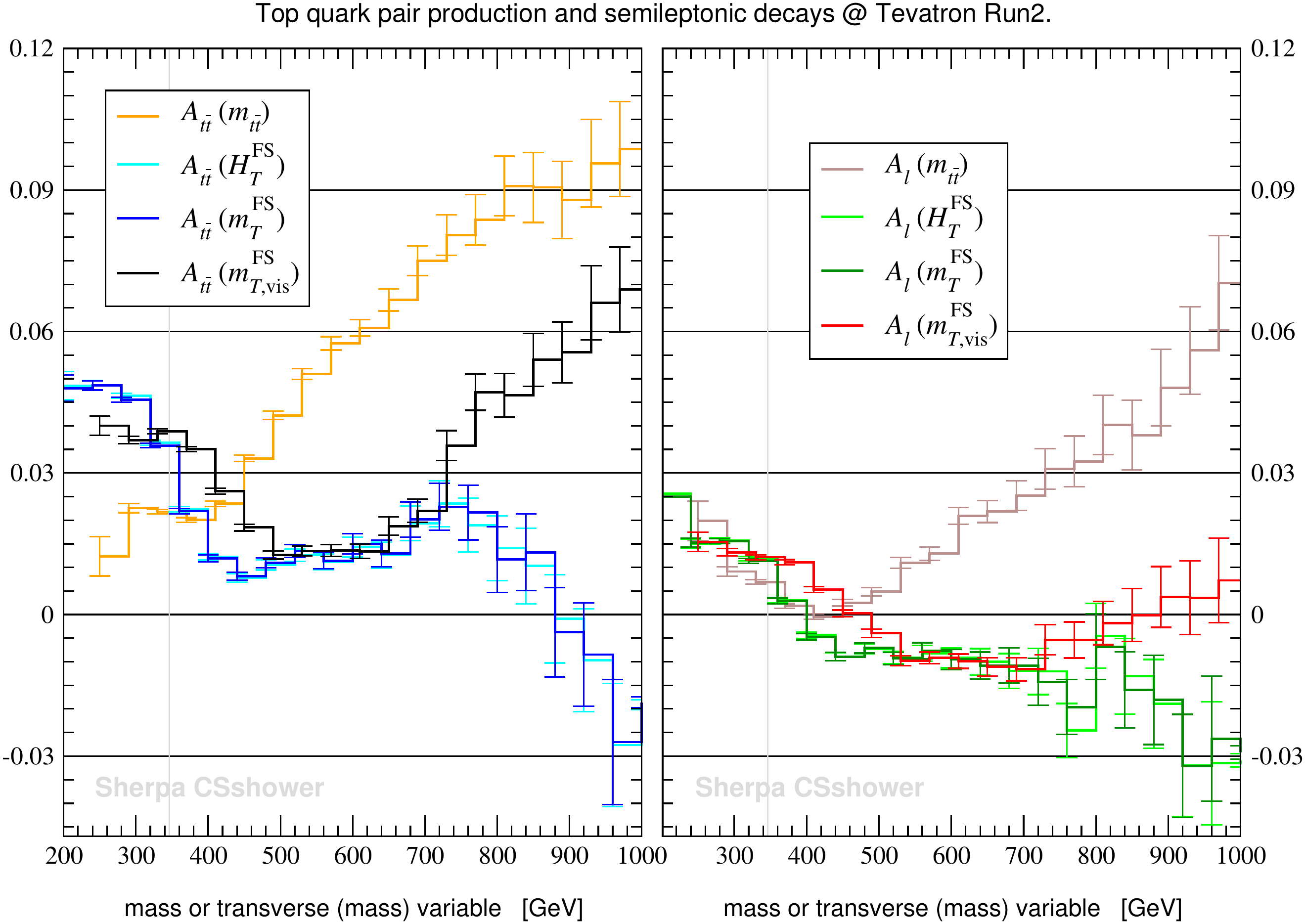}
\caption{\label{fig:sherpa.extra}
  Differential forward--backward (left) and lepton-based asymmetries
  (right) in $t\bar t$\/ events as a function of various transverse
  observables (and for comparison, as a function of $m_{t\bar t}$)
  using CDF-like event selection (see Sec.~\ref{sec:cuts}). The
  results in the top two panels were obtained by running \powheg +
  \pythia, while the bottom panels show results from \sherpa (all
  generated as in Sec.~\ref{sec:reco}).}
\end{figure}

Studies using reconstruction-independent asymmetries and/or variables
such as these are useful for a couple of reasons. First,
reconstruction-independent quantities have no (intrinsic)
combinatorial issues, so the sensitivity to how these objects are
chosen is significantly reduced. Second, as they are inclusive, these
quantities are less sensitive to the modeling of the various
distributions involved in the actual measurements. As a result,
systematic uncertainties are under better control. Third, we can
introduce additional analysis levels with varying degree of
reconstruction-dependence (\eg reconstruction-dependent asymmetry,
$\At$, versus reconstruction-independent observable, $m_T$), allowing
for a number of sanity and closure tests on the existing data.
For example, if the Tevatron
collaborations were to measure the dependence of the asymmetries on
one of the variables introduced above and find agreement with our
studies, that would strongly hint at an error in the reconstruction.
With no more data-taking at the Tevatron it is essential to try out
each different angle in forming differential asymmetry measurements as
it might be the only way to shed more light on the present data versus
theory puzzle.

To illustrate this in more detail, we present in \Fig{fig:sherpa.extra}
the differential top quark (left panels) and lepton (right panels)
asymmetries with respect to the final-state variables discussed above. These
events were generated using \powheg + \pythia (top panels), and
simultaneously with \sherpa (bottom panels) following the same
procedure as in \Sec{sec:reco}. For comparison, we also show the
differential asymmetries with respect to $m_{\ttbar}$. The qualitative
similarity between results in the different generators is evidence for the
decreased modeling sensitivity mentioned above.

Focusing on $\At$, the asymmetry with respect to $H_T$ does not rise
nearly as fast as $\At\,(m_{\ttbar})$. This occurs because $H_T$ is
sensitive to extra radiation in the event, while $m_{\ttbar}$ is
not. Events that have large $H_T$ due to lots of extra radiation will
carry a large, negative asymmetry, partially counteracting the large
positive asymmetry generated by large $m_{\ttbar}$, large $H_T$
events. Similarly, the design of the selection cuts is such that in
most cases the visible
transverse momentum is balanced out by the missing transverse energy
vector. This yields an asymmetry dependence on $m_T$ that is very close
to that of $H_T$.  Finally, we use $m_{T,\mrm{vis}}$ rather than $m_\mrm{vis}$
alone, because $m_{T,\mrm{vis}}$ contains some missing energy
information and thereby gives a better approximation to the total
invariant mass of the $t\bar t\,$+$X$\/ system. The effect only kicks in
for larger transverse masses where $\At\,(m_{T,\mrm{vis}})$ starts to
follow the $m_{t\bar t}$ dependence of $\At$. In the low $m_{T,\mrm{vis}}$ region
the transverse character of the observable still dominates, leading to
an $H_T$-like asymmetry dependence.

Turning to the dependence of $\Al$, we find that the differential
leptonic asymmetry actually is quite insensitive to the choice of the
specific transverse variables and is very small in magnitude
($|\Al(x)|\le\text{2\%}$), potentially too small to be observed at the
Tevatron (even without requiring top quark reconstruction). However,
this argument can be turned around in the sense that the standard
model gives predictions that by and large are ``consistent with
zero''. Hence any measurement deviating from zero in these observables
can be considered as a fairly clear signal for new physics.

It is interesting to explore these ``reconstruction-free''
distributions both from an experimental point of view as well as from
the theoretical perspective, where one should study the robustness of
this new set of observables more carefully. Such a semi-inclusive
approach can, in principle, lead to a cleaner set of precision
observables and, in the future, also turn out to be useful for the LHC
experiments.

\section{Conclusions}\label{sec:conclusions}

The $\ttbar$ asymmetry measured by the CDF and D\O\ collaborations remains higher than expected in the standard model (SM), both when measured inclusively and when binned in kinematic variables such as $m_{\ttbar}$ or $Y_{\ttbar}$. In order to determine the origin of the discrepancy, in this paper we have proposed that the 
leptonic asymmetry $\Al$ is a useful probe. Within the SM, $\Al$ is
inherited from $\At$, with the relationship between the two set by the
kinematics of the $\ttbar$ system. Thus, by varying the lepton $p_T$
-- a simple proxy for the mother-top quark energy, directionality and
spin --  one obtains a simple correlation between the top quark and lepton-based asymmetries which can be conveniently described by a curve $\Al[\Atp]$ in the $\Al-\At$ plane. We have verified that this correlation between asymmetries is qualitatively maintained through all levels of the analysis, from parton level through the inclusion of showering and reconstruction. The correlation is also stable under variations of theory inputs and under potential mis-modeling. 

By studying the full $\Al-\At$ correlation in data (and their
individual $\ptl$ dependence), rather than just the inclusive
asymmetries, we gain discriminating power.
The slope of the curve is insensitive to the size of the overall
normalization of the asymmetry and is found to be rather stable
against various deformation of inputs and analysis parameters. The
slope is, however, not satisfactorily modeled by LO + PS tools and
requires complete NLO (+ PS) treatment.
We have demonstrated the discriminating power of the
$\Al\,(\ptl)[\At\,(\ptl)]$ correlation by showing several beyond the
SM benchmarks with the same net asymmetry but with correlations in the
$\Al-\At$ plane that are all distinct and distinct from the SM
case. This argument for studying the correlation among asymmetries is
not unique to the semileptonic $\ttbar$\/ channel, nor is it unique to
the Tevatron. It is important to keep in mind that, even if the
central values of observables such as $A_C$ at the LHC turn out to
match the SM values within uncertainties, the correlation of $A_C$
with other asymmetries in the $\ttbar$\/ system are a worthwhile test
of the SM, and can be proven to be more stable similar to the case
discussed above. We have also briefly considered how the asymmetry can
be measured via more inclusive, ``reconstruction-free''
variables. These can be further used in principle to control the
measurement of top quark precision observables and might provide us
with extra sensitivity to the presence of non-standard model
contributions to the various asymmetries.

% ==============================================================================

\section*{Acknowledgments}

We are thankful to Dan Amidei, Keith Hamilton, Amnon Harel, Stefan
H\"oche, Paolo Nason, Gavin Salam, Michael Trott and Ciaran Williams
for many helpful discussions. GP is the Shlomo and Michla Tomarin
development chair, supported by the grants from GIF, Gruber
foundation, IRG, ISF and Minerva.

% ======= bibliography =========================================================

\bibliographystyle{JHEP}%{amsunsrt_mod}
{\raggedright\frenchspacing\bibliography{FBAlAtt}}
 
\vspace*{0mm}\noindent\hrulefill

% ======= the end ==============================================================

%\end{fmffile}
\end{document}

%% file: texsetup.tex
\usepackage{amsmath}
\usepackage{amssymb}
\usepackage{latexsym}
\usepackage{amsfonts}
\usepackage[mathscr]{eucal}
\usepackage{array}
\usepackage{calc}
\usepackage{longtable}
\usepackage{multirow}
\usepackage{bm}% bold math
\usepackage{fixmath}
\usepackage{upgreek}
\usepackage{isomath}
\usepackage{nicefrac}
\usepackage{slashed}
\usepackage[usenames,dvipsnames,svgnames]{xcolor}
\usepackage{graphicx,psfrag}
\usepackage{axodraw}
\usepackage{xspace}
\usepackage{subfigure}
\usepackage[space,nosort]{cite} % use compact form later

\usepackage{setspace}
\usepackage{bookman}
\usepackage{chancery}
\usepackage{lmodern}
\usepackage[OT1,T1]{fontenc}
\usepackage{textcomp}
\usepackage[font=small,labelfont=bf]{caption}%format=hang
\usepackage{sectsty}
\allsectionsfont{\sffamily}
\subsubsectionfont{\bfseries\slshape\sffamily}
\DeclareMathAlphabet{\mathitbf}{OML}{cmm}{b}{it}
\DeclareMathAlphabet{\mathbook}{T1}{pbk}{m}{it}   %take from ``bookman''
\DeclareMathAlphabet{\mathzapf}{T1}{pzc}{m}{n}    %take from ``chancery''
\usepackage{fancyhdr}
\let\spreprint\empty

\let\sinstitute\empty

\makeatletter
\newcommand{\mymaketitle}{\begingroup
  \null\thispagestyle{empty}%
    \ifx\spreprint\empty
      \vskip5ex
    \else
      \vspace*{-10ex}\flushright\small\spreprint\vskip2ex
    \fi
    \vskip7ex
    \begin{center}
      \doublespacing{\sffamily\bfseries\LARGE\@title}\vskip5ex
      {\bf\large\@author\vskip 2ex}
      \ifx\sinstitute\empty
      \else
        {\sl\small\sinstitute}
      \fi
    \end{center}
    \vskip7ex
  \endgroup
}
\makeatother
\renewenvironment{abstract}{\begin{center}
  {\large\sffamily\bfseries Abstract}\\[4mm]
  \begin{minipage}[t]{0.87\textwidth}\small
}{\end{minipage}\end{center}\vskip10ex}

\usepackage[pdfborder={0 0 0}]{hyperref}
\hypersetup{% title % author % list of keywords
  pdftitle={Data driving the top quark forward--backward asymmetry
    with a lepton-based handle},
  pdfauthor={A.Falkowski, M.L.Mangano, A.Martin, G.Perez, J.Winter},
  pdfkeywords={Hadron Colliders}{Top Quarks}{Asymmetry},
  colorlinks=true,         % false: boxed links; true: colored links
  linkcolor=MediumBlue,      % myblue, red, color of internal links
  citecolor=DodgerBlue,    % mydarkgreen, green, color of links to bibliography
  filecolor=FireBrick,     % darkred, magenta, color of file links
  urlcolor=DarkGreen       % darkgreen, cyan, color of external links
}
\usepackage[all]{hypcap}     % good, but check to what extent figures jump if %
\newcommand{\bit}{\begin{itemize}}
\newcommand{\eit}{\end{itemize}}
\newcommand{\bc}{\begin{center}}
\newcommand{\ec}{\end{center}}
\newcommand{\bq}{\begin{quote}}
\newcommand{\eq}{\end{quote}}
\newcommand{\bi}{\begin{itemize}}
\newcommand{\ei}{\end{itemize}}
\newcommand{\beq}{\begin{equation}}
\newcommand{\eeq}{\end{equation}}
\newcommand{\bea}{\begin{eqnarray}}
\newcommand{\eea}{\end{eqnarray}}
\newcommand{\bt}{\begin{tabular}}
\newcommand{\et}{\end{tabular}}
\newcommand{\bmp}{\begin{minipage}}
\newcommand{\emp}{\end{minipage}}
\newcommand{\btab}{\begin{tabbing}}
\newcommand{\etab}{\end{tabbing}}

\newcommand{\pythia}{P\protect\scalebox{0.8}{YTHIA}\xspace}
\newcommand{\sherpa}{S\protect\scalebox{0.8}{HERPA}\xspace}
\newcommand{\mcfm}{MC\protect\scalebox{0.8}{FM}\xspace}
\newcommand{\powheg}{P\protect\scalebox{0.8}{OWHEG}\xspace}

\newcommand{\cssr}{CS\protect\scalebox{0.8}{SHOWER}\xspace}

\newcommand{\fjet}{F\protect\scalebox{0.8}{AST}J%
\protect\scalebox{0.8}{ET}\xspace}
\def\trm{\textrm}
\def\nn{\nonumber}

\def\mrm{\mathrm}
\def\cal{\mathcal}

\def\eg{\textsl{e.g.\xspace}}
\def\ie{\textsl{i.e.\xspace}}

\def\eq#1{eq.~(\ref{#1})}

\def\Fig#1{Figure~\ref{#1}}

\def\Sec#1{Section~\ref{#1}}